\documentclass[preprint,journal]{vgtc}            % preprint (journal style)

\onlineid{0}

\vgtccategory{Research}

\title{Who’s That Player?: Externalizing Query Interpretation in Spoken XR Sports Interaction}

\author{%
    \authororcid{Chunggi Lee}{0000-0002-6164-2563},
  \authororcid{Tica Lin}{0000-0002-2860-0871},
  \authororcid{Yalong Yang}{0000-0001-9414-9911}, and
  \authororcid{Hanspeter Pfister}{0000-0002-3620-2582}
}

\authorfooter{
  \item
  	Chunggi Lee and Hanspeter Pfister are with Harvard University.
  	E-mail: \{chunggi\_lee, pfister\}@g.harvard.edu.
  \item
  	Tica Lin is with Dolby Laboratories.
  	E-mail: mlin@g.harvard.edu.

  \item Yalong Yang is with Georgia Tech.
  	E-mail: yalong.yang@gatech.edu.
}

\abstract{%
  XR sports viewing enables spectators to follow play from immersive, spatially anchored perspectives while accessing contextual analytics directly within the scene. In such settings, speech offers a practical interaction modality because text entry and menu navigation can interrupt attention during fast-paced gameplay. However, spoken queries are often underspecified: viewers may omit which player, time period, field location, or metric they intend. When systems resolve these ambiguities implicitly, their assumptions remain hidden, making misinterpretations difficult to notice and correct (repair). We investigate how externalizing a system's interpretation of spoken queries can support inspection and correction of such misunderstandings in XR sports viewing. Through a formative study, we identified four recurring ambiguity types (referential, spatial, temporal, and metric) that characterize ambiguous spoken queries in this context. We develop a design space that organizes externalization along three dimensions (ambiguity type, interpretation state, externalization strategy) and instantiate it in an interactive XR soccer viewing system that combines situated visual cues with supporting analytic views. A within-subjects user study (N=16) comparing externalized interpretation against a voice-only baseline reveals that externalization is associated with higher inspectability on most measured dimensions and increased explicit repair language overall. However, repair occurred in only 38\% of misaligned externalization trials, and this visibility-action gap varied by ambiguity type, indicating that transparency and correction affordance are orthogonal design axes.
}


\teaser{
  \centering
  \includegraphics[width=\linewidth, alt={A view of a city with buildings peeking out of the clouds.}]{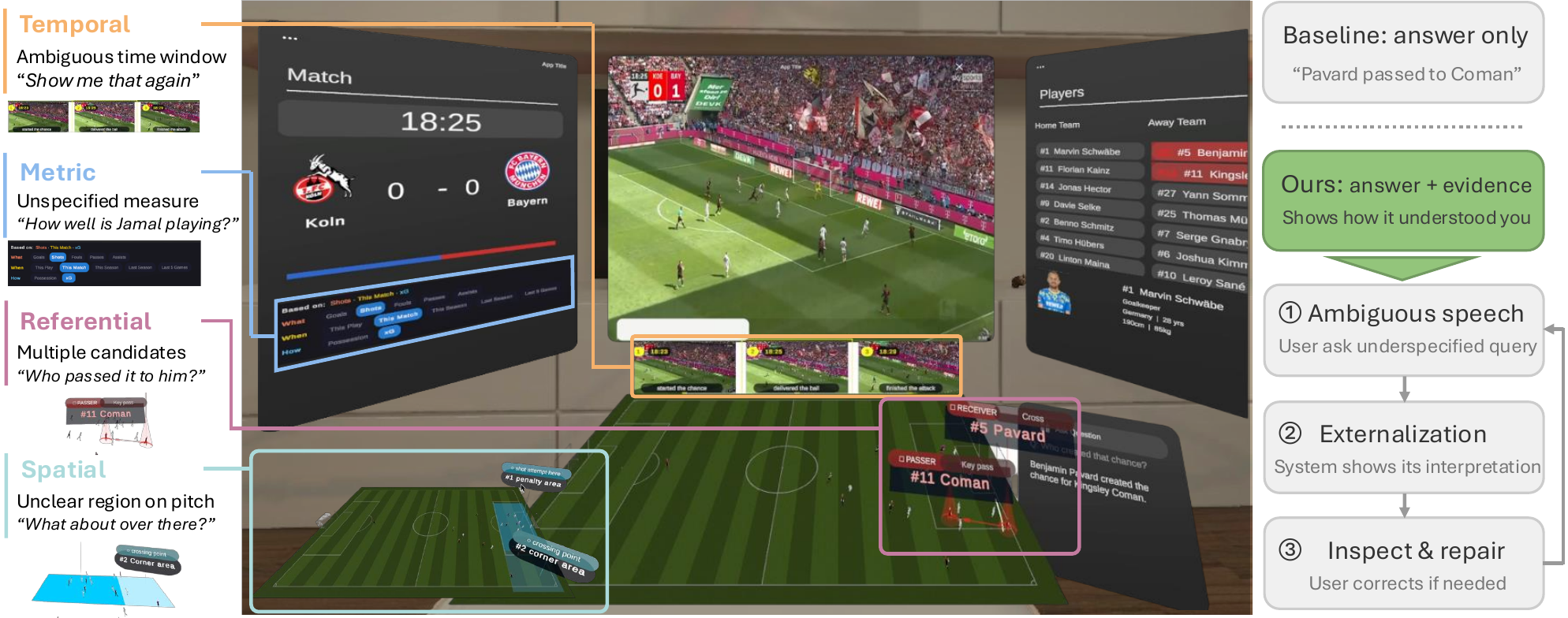}
  \caption{%
    Our XR soccer viewing system externalizes the system's interpretation of ambiguous spoken queries. (Left) Four ambiguity types, each visualized as situated cues within the scene. (Center) The system externalizes its interpretation through spatially anchored visual cues and analytic views. (Right) Users can inspect system assumptions and repair mismatched interpretations.
  }
  \label{fig:teaser}
}

\graphicspath{{figs/}{figures/}{pictures/}{images/}{./}} % where to search for the images

\usepackage{booktabs}                  % only used for the table example
\usepackage{lipsum}                    % used to generate placeholder text
\usepackage{mwe}                       % used to generate placeholder figures
\usepackage{ccicons}                   % package to be able to use icons from creative commons

\usepackage{mathptmx}                  % use matching math font

\usepackage{tabu}                      % 
\usepackage{multirow}

\usepackage{dblfloatfix}
\usepackage{amsmath}

\usepackage[utf8]{inputenc}
\usepackage{listings}
\usepackage{xcolor}
\usepackage{booktabs}

\newcommand{\rev}[1]{\textcolor{black}{#1}}

\begin{document}

\newcommand{\cglee}[1]{\textcolor{orange}{[CG: #1]}}
\newcommand{\tl}[1]{\textcolor{orange}{[TL: #1]}}

%%%%%%%%%%%%%%%%%%%%%%%%%%%%%%%%%%%%%%%%%%%%%%%%%%%%%%%%%%%%%%%%
%%%%%%%%%%%%%%%%%%%%%% START OF THE PAPER %%%%%%%%%%%%%%%%%%%%%%
%%%%%%%%%%%%%%%%%%%%%%%%%%%%%%%%%%%%%%%%%%%%%%%%%%%%%%%%%%%%%%%%


\firstsection{Introduction}
\maketitle

% \cglee{aligned vs misaligned / emb vis --> externalizing / live play or others / target users etc / system remake / flow / result figure / playing format - keep going or replay}

% 1. aligned vs misaligned
% need not to mention aligned and misaligned  in introduction - more simple one 
% justify and make sure - balanced - how people react incorrect responses / add explanation what people or other papers did

% 2. novelty: sport xr and other things are new
% good idea - some existing design space and integrate together
% Tica, AI papers, and sports - better way to say novelty - adding references and explanation for design spaces

% 3. embedded vis: XAI - ivuslaization to explanation - using vis - interpretabilty - introduction need to be clear, existing approach not very visual or just multiple choices and not enough, very high spatial temporal ones, we need a real visualization process.

% 4. Interpretation state need to write - not straightforward to write - terminology from XAI / one sentence to explain each individual part in design space

% 5. not live sports - motivated by live watching, want to focus live  / sport viewers or watchers / different level of expreince different types of questions - benefits of llm and robust 

% 6. open a big and narrow down. it's big topic, we focus on xai in xr or sports

Sports viewing is one of the most pervasive forms of media consumption worldwide~\cite{giorgio2023sports}. As advances in sensing and tracking technologies have made fine-grained match data increasingly available~\cite{bassek2025integrated}, growing research interest has turned to designing richer, more analytic viewing experiences that go beyond passive spectatorship~\cite{lee2024sportify, zhi2019gameviews, chen2021viscommentator}. XR sports viewing enables spectators to follow play from immersive, spatially anchored perspectives while accessing contextual analytics directly within the scene~\cite{lin2023vird, lin2022omnioculars, lo2021xrspectator}, where player statistics, match data, and multiple analytic views can be spatially anchored within the three-dimensional environment. In such settings, speech offers a practical interaction modality: mid-air text entry is slow and error-prone~\cite{knierim2020opportunities, kern2023text}, sustained gestural interaction can cause physical fatigue~\cite{hincapie2014consumed, jang2017modeling}, and both alternatives require viewers to divert visual attention from the ongoing game. As natural language interaction gains traction in XR environments through LLM-powered agents~\cite{buldu2025cuify, tang2025llmxr}, speech is becoming an increasingly viable modality for accessing contextual information during immersive viewing.

However, spoken queries in this context are often imprecise. Sports viewing is fast-paced and cognitively demanding: viewers must track multiple players, follow evolving play, and react to events, leaving limited capacity to formulate detailed queries. Unlike formal analytic settings where users can deliberate over query construction, viewers tend to produce short, context-dependent utterances that omit which player, time period, field location, or metric they intend, asking questions like \textit{``Who's that player?''} or \textit{``Show me that play again.''} 
While ambiguity is a property of natural language~\cite{ min2020ambigqa, setlur2019inferencing}, interpretation in situated immersive environments depends strongly on visual context, spatial reference, and conversational history~\cite{song2026sia}, making sports viewing a particularly demanding case: games involve multiple moving entities, rapidly changing events, and frequent shifts in attention.

Multimodal approaches (e.g., combining speech with eye gaze~\cite{lee2024gazepointar} or pointing~\cite{lee2021whatsthis}) can help disambiguate references, but offer limited support when multiple entities overlap spatially or when context shifts rapidly. Even with such additional modalities, aspects of query ambiguity beyond referent resolution, such as temporal scope, analytic framing, and comparison targets, often remain unspecified~\cite{setlur2019inferencing}. When systems resolve these ambiguities implicitly, their assumptions remain hidden, and users receive only the final answer without visibility into the interpretation process~\cite{gao2015datatone, liu2023afraid}.
Prior work on ambiguous speech in interactive systems has primarily emphasized disambiguation accuracy, using context, multimodal signals, or clarification to infer user intent~\cite{lee2024gazepointar, lee2021whatsthis, song2026sia, kiesel2018voice}. Less attention has been given to how the system's chosen interpretation is exposed to users once it commits to one~\cite{liu2023afraid, gao2015datatone}. In XR sports viewing, this distinction is consequential: viewers must judge whether the system understood them correctly while play continues and attention remains on the scene. 
We therefore treat ambiguity not only as a system-side inference problem, but as a \textit{user-facing interaction problem} in which a system's inferred referent, temporal framing, or analytical scope must be made available for inspection and correction.
XR sports viewing remains particularly challenging in this regard because rapidly changing game states, overlapping entities, and divided viewer attention make it difficult both for systems to resolve ambiguity correctly and for users to notice when they have not.

\begin{figure}[t]
\centering
\includegraphics[width=\linewidth]{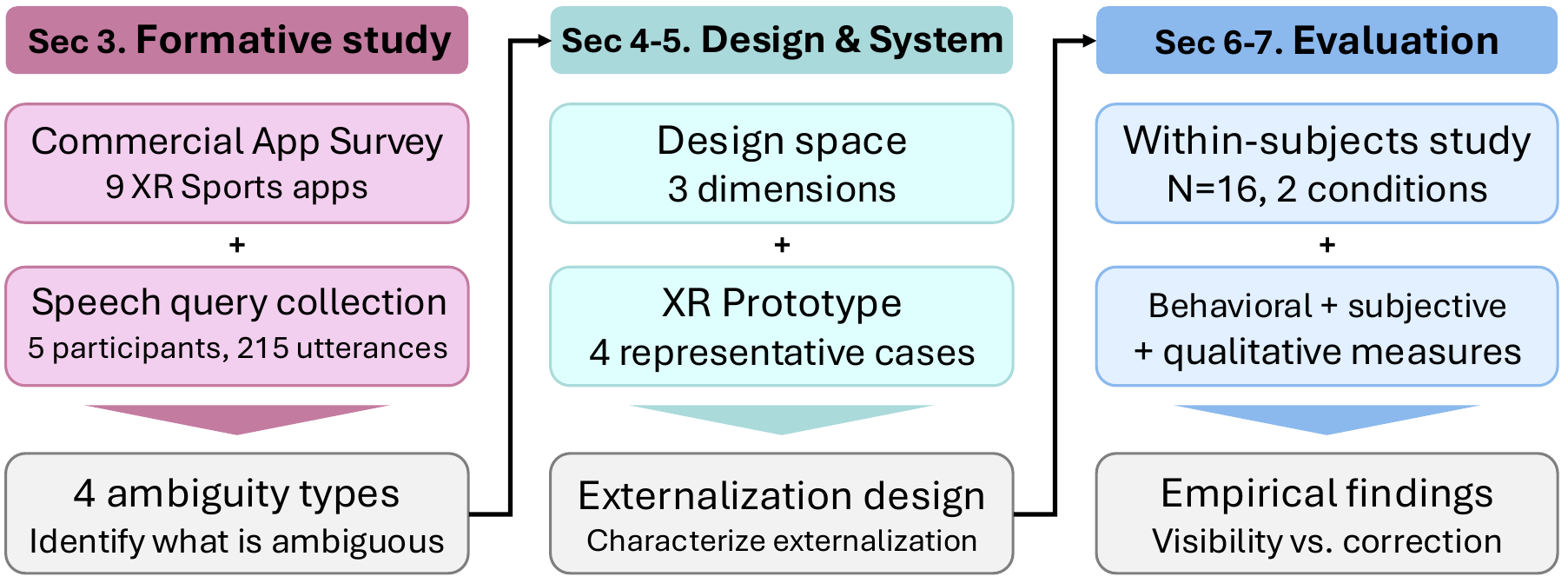}
\caption{Overview of our three-stage research process. Ambiguity types identified in the formative study informed the design space and system, which were then evaluated in a controlled user study.}
\label{fig:research_process}
\vspace{-1em}
\end{figure}

% We investigate how \textbf{externalizing} a system's interpretation of ambiguous spoken queries can support such inspection and correction in XR sports viewing. Drawing on prior work in speech disambiguation and interpretation transparency, we aim to address the underexplored problem of making system-inferred interpretations of spoken queries visible and correctable.
% We define externalization as making the system's interpretation process visible within the viewing environment: showing which entity the system selected, what time window or metric it assumed, and what alternatives it considered, so that users can verify the system's understanding and correct (repair) it when needed. 
\rev{We investigate how \textbf{externalizing} a system's interpretation of ambiguous spoken queries can support inspection and correction in XR sports viewing. Rather than treating externalization as a generic explanation panel or candidate list, we define it as a situated visualization process: representing the system's inferred player, field region, time window, metric, or comparison target in relation to the reconstructed match scene and associated analytic views, so that viewers can verify and repair the system's understanding while play continues.}
Our research proceeds in three stages (Fig.~\ref{fig:research_process}). First, through a formative study in which participants watched soccer clips and spoke questions aloud, we identified four recurring ambiguity types \textbf{(referential, spatial, temporal, and metric)} that characterize what remains underspecified in spoken queries. Second, we developed a design space for externalizing query interpretations and instantiated it in an interactive XR soccer viewing system combining situated visual cues with analytic views across four cases (identity resolution, spatial clarification, performance comparison, and moment retrieval). Third, we evaluated the system through a within-subjects user study ($N=16$) comparing externalized interpretation against speech-only interaction. Using reconstructed match data~\cite{lin2022omnioculars, lee2024sportify, chen2023iball}, participants viewed reconstructed plays across both correctly interpreted and deliberately misaligned trials.

% Our results show that externalization was associated with higher inspectability on three of four measured items without imposing additional interaction cost, and participants produced explicit repair language at a higher overall rate. 
\rev{Our empirical finding is a visibility-action gap: externalization made system assumptions more inspectable, but visible assumptions did not reliably translate into corrective action.}
To examine repair behavior under controlled conditions, we included both aligned trials (system interpretation matches intent) and deliberately misaligned trials (shifted to a plausible but non-primary reading), ensuring that misaligned interpretations were reliably present across all four ambiguity types. Notably, in aligned trials, referential externalization was accompanied by active verification rather than passive acceptance, consistent with grounding theory that visible commitments invite ratification~\cite{clark1991grounding}. At the same time, we observed a 62\% gap between externalized visibility and corrective action in misaligned trials: the system surfaced its assumptions, but participants often did not correct them, particularly when repair required precise verbal specification. These findings suggest that making system assumptions visible is necessary but not sufficient for effective correction, and that future designs should pair externalization with lower-cost correction pathways to close the gap between noticing and acting.

% This work makes three contributions. First, we frame ambiguity in speech-driven XR sports viewing as a user-facing interaction problem, grounded in a formative study identifying four recurring ambiguity types. 
% Second, we develop a design space organized along three dimensions (ambiguity type, interpretation state, externalization strategy) and instantiate it in an XR soccer viewing system across four cases. 

\rev{First, we frame ambiguity in speech-driven XR sports viewing as a situated interpretation problem, grounded in a formative study identifying four recurring ambiguity types. 
Second, we develop a design space for externalizing interpretation commitments through visual and analytic representations, organized along three dimensions (ambiguity type, interpretation state, externalization strategy), and instantiate it in an XR soccer viewing system across four cases.}
Third, we provide empirical evidence that transparency and correction affordance are orthogonal design axes: \rev{externalization supported higher inspectability and more repair language than the voice-only baseline}, yet repair occurred in only 38\% of misaligned trials, with the gap varying by ambiguity type in ways that indicate correction cost, not transparency quality, as the binding constraint.

\section{Related Work}

% \subsection{Sports Spectating, Question Answering, and Analytic Systems}
\subsection{Interactive Sports Spectating and Analytics}
Prior work has explored interactive systems for sports spectating, question answering, and analytic support across conversational, video, and immersive settings. For example, GameBot supports conversational access to sports statistics~\cite{zhi2020gamebot} and GameViews presents in-game box scores through interactive visualizations~\cite{zhi2019gameviews}. XRSpectator augments immersive spectating with situated visual information~\cite{lo2021xrspectator}, ARSpectator explores mobile AR overlays for sports events~\cite{zollmann2019arspectator}, and Stats on-site proposes a framework for situated visualization approaches for on-site sports spectating~\cite{zollmann2021stats}. More recent systems embed visualizations directly into sports video: iBall uses gaze-moderated visualizations to reduce visual clutter in basketball~\cite{chen2023iball}, and VisCommentator augments sports videos with automated visual commentary~\cite{chen2021viscommentator}. Sportify combines question answering with embedded visualizations and personified narratives~\cite{lee2024sportify}, and AiCommentator integrates conversational interaction with embedded visualization for football broadcasts~\cite{andrews2024aicommentator}.

% Beyond spectating, immersive analytics systems support coaching and tactical analysis: TiVee enables badminton tactics exploration through immersive 3D visualizations~\cite{chu2022tivee}, VIRD reconstructs match videos for high-performance coaching~\cite{lin2023vird}, VisCourt provides in-situ tactic guidance in mixed reality~\cite{cheng2024viscourt}, and Smartboard and Team-Scouter support LLM-driven tactic exploration and player scouting for soccer~\cite{liu2025smartboard, cao2025teamscouter}. Targeting sports analysts, BKViz provides interactive visual analytics for basketball player and team dynamics~\cite{losada2016bkviz}, OBTracker evaluates off-ball movements through novel glyph designs~\cite{wu2022obtracker}, and HoopInSight compares shooting performances using spatial visualizations~\cite{fu2023hoopinsight}. Related efforts have examined injury risk~\cite{lee2025vair} and embodiment transformation for inclusive sports experiences~\cite{saiki2026bridge}. 
Beyond spectating-focused systems, broader efforts in immersive sports analytics have explored coaching and tactical analysis~\cite{chu2022tivee, lin2023vird, cheng2024viscourt}, LLM-driven tactic exploration and player scouting~\cite{liu2025smartboard, cao2025teamscouter}, visual analytics for player and team evaluation~\cite{losada2016bkviz, wu2022obtracker, fu2023hoopinsight}, as well as injury risk assessment and inclusive sports experiences~\cite{lee2025vair, saiki2026bridge}. While these systems demonstrate the value of embedding analytic information within sports contexts, they rely primarily on menu-driven or direct manipulation interfaces. With the growing adoption of LLM-based agents in XR environments~\cite{tang2025llmxr, buldu2025cuify}, natural language interaction is becoming an increasingly prominent modality for accessing sports data, as demonstrated by conversational and question-answering sports systems~\cite{lee2024sportify, andrews2024aicommentator}. However, the user experience challenges that arise when spoken queries are ambiguous or underspecified remain largely unaddressed in sports viewing. Our work builds on this setting but focuses on ambiguity in spoken interaction during XR sports viewing, examining how underspecified queries leave reference, time, or comparison intent unresolved, and how the resulting system interpretation can be made inspectable.

\subsection{Ambiguity and Grounding in Speech-Based Interaction}
Ambiguity is a recurring property of speech-based interaction, especially when users produce short or underspecified utterances, where the intended referent, temporal scope, or comparison target may remain unclear and multiple interpretations may be plausible for the same utterance~\cite{clark1991grounding}. Prior work has studied this problem from both interaction and modeling perspectives, including voice query clarification~\cite{kiesel2018voice}, underspecified language in interactive analysis~\cite{setlur2019inferencing}, and ambiguous open-domain questions that admit multiple plausible answers~\cite{gao2021answering, min2020ambigqa}. In the context of natural language interfaces for data visualization, systems such as DataTone~\cite{gao2015datatone} and XNLI~\cite{shen2024xnli} have shown that making system decisions visible can support user correction behavior, while Liu et al.~\cite{liu2023afraid} show that LLMs often commit to a single interpretation without acknowledging alternatives. These challenges are compounded in situated environments where interpretation depends on visual context, spatial reference, and conversational history.

Prior approaches address ambiguity through mediation~\cite{dey2005mediation}, mixed-initiative interfaces~\cite{horvitz1999mixed}, clarification dialogues~\cite{gao2015datatone, ma2025ambigchat}, and multimodal grounding via gaze or pointing~\cite{lee2021whatsthis, lee2024gazepointar}. However, even when additional modalities are available, ambiguity may still remain: gaze may indicate a region rather than a single player, and pointing may underspecify the intended target among nearby or overlapped entities. Most prior work focuses on preventing or resolving ambiguity before the system acts~\cite{song2026sia}, which is effective when user intent can be anticipated. In fast-paced viewing contexts, however, users often produce spontaneous queries whose ambiguity only becomes apparent after the system responds. Our work addresses what happens after the system commits to an interpretation, making that interpretation visible so users can inspect and correct it in a context where rapidly changing events and shifting attention make spoken queries especially context-dependent.

\subsection{Explainable AI and Interpretation Transparency}
Transparent and interpretable AI research emphasizes making system reasoning visible to users~\cite{hoffman2023measures, liao2023designerly, chromik2021human}. Chromik and Butz~\cite{chromik2021human} argue that effective explanations must be tailored to the user's task context rather than presented as generic outputs. Findings on whether transparency improves human-AI collaboration are mixed: explanations do not consistently improve decision accuracy~\cite{lai2019human} or team performance~\cite{bansal2021whole}, cognitive forcing functions can reduce overreliance~\cite{bucinca2021trust}, and visible AI outputs can anchor users to initial interpretations even when errors are present~\cite{nourani2021anchoring}. Vasconcelos et al.~\cite{vasconcelos2023explanations} show that explanations can reduce overreliance under certain conditions, suggesting the effect depends on presentation context.

Recent work has begun translating these insights into AR and XR. Xu et al.~\cite{xu2023xair} propose XAIR, a framework addressing when, what, and how to present AI explanations in AR, and Maathuis et al.~\cite{maathuis2025xaixr} survey the broader integration of XAI methods in XR systems, identifying transparency of AI reasoning as a key open challenge across application domains. However, both focus primarily on explaining why a system produced a given output, such as why an item was recommended or an object was recognized. In speech-driven sports viewing, the transparency challenge is different: users need to see not why the system chose a response, but how it interpreted their query, including which entity it selected, what time window it assumed, and what scope it inferred. We examine how users inspect and act on such interpretation cues embedded within an XR viewing context.
\section{Formative Study}
% To inform our system design, we conducted a two-part formative investigation: (1) a survey of commercially available XR sports viewing applications to understand the current landscape and (2) a study in which five participants watched soccer clips and spoke questions aloud. Following prior formative-study-driven HCI work, we organize this section into study setup and findings~\cite{lee2026vistar, lin2022omnioculars}.

\rev{To inform our system design, we first reviewed commercially available XR sports-viewing applications as contextual motivation, then conducted a formative study in which five participants watched soccer clips and spoke questions aloud.} Following prior formative-study-driven HCI work, we organize this section into study setup and findings~\cite{lee2026vistar, lin2022omnioculars}.

\subsection{XR Sports Viewing App Layout and Interaction Analysis}
\label{sec:apps}
\rev{To understand what interaction modalities and information layouts current XR sports viewing products offer, we surveyed nine commercial applications across Apple Vision Pro, Meta Quest, and web-based platforms, covering seven sports (see supplementary material for details).
We selected commercially available XR/immersive sports-viewing applications across major platforms and sports, and coded public materials for layout, supporting panels, alternate views, tabletop content, and input modalities.}
Across all applications, a consistent layout emerged: a central main video view flanked by supporting panels for statistics, player profiles, rankings, or alternate camera angles, with some apps adding a tabletop surface for secondary content (e.g., ESPN for basketball).
However, the information in these panels is pre-configured or menu-driven, limiting users to browsing fixed categories rather than requesting what they need in the moment. Speech offers a practical modality for such on-demand querying during immersive viewing, yet none of the surveyed applications support it. To understand how viewers formulate spoken queries in XR sports viewing and what challenges arise for system interpretation, we conducted a formative study with five participants.

\subsection{Study Setup}

\textbf{Participants.}
We recruited five participants ($N$=5: 4 male, 1 female), all aged between 25 and 34 years old.
Participants reported diverse levels of soccer engagement: viewing frequency ranged from a few times a year to very frequently, and they rated their soccer knowledge relatively high (\(M=5.4\), \(SD=0.9\) on a 7-point scale).
This range reflects our target population of casual to moderately engaged viewers.
Two participants had prior experience with AR/VR devices (e.g., Meta Quest, Apple Vision Pro), while three had not.
The study was approved by Harvard University’s IRB (Protocol No. IRB25-0906), and all participants provided informed consent.
% The study protocol was reviewed and approved by the institutional review board at Harvard University (Protocol No. IRB25-0906). All participants provided informed consent prior to participation.

\textbf{Setup.}
Following the spatial layout patterns observed in Sec~\ref{sec:apps}, we set up a viewing environment with a main video feed in the center, a tabletop display below for the game overview, and information panels on either side for game and player details(~\autoref{fig:teaser}). For this formative study, we used only the basic voice pipeline (speech-to-text, context builder, and LLM response generation in Figure~\ref{fig:system}), without the ambiguity handler or visual externalization, which were developed based on findings from this study. Participants interacted via push-to-talk: pressing and holding the ``B'' button on the controller initiated recording, and releasing it triggered speech-to-text conversion. This ensured reliable segmentation of individual utterances and avoided latency issues from continuous recognition. All transcribed utterances were logged.

\textbf{Procedure.} Each session lasted 40--60 minutes and consisted of three phases: a demographic pre-survey, a video-viewing and question-asking task, and a follow-up interview. In the main task, participants watched five short clips ($\sim$30s each), each depicting a distinct match situation (e.g., a goal, a foul, a set piece). To keep questions relevant to match context, we structured the elicitation around common question categories from prior sports-viewing research, such as identifying players, retrieving statistics, comparing performance, and replaying moments~\cite{lin2022omnioculars, chen2023iball, lee2024sportify, giorgio2023sports}. For each clip, we provided two scenario descriptions (e.g., \textit{``You're curious how the goal scorer's performance compares to the other attackers.''}), followed by free questioning. Participants then watched a longer segment ($\sim$4 min) of continuous play and asked questions freely throughout. We collected 215 utterances (M=43 per participant).

\textbf{Data Analysis.}
% We reviewed and cleaned the transcribed utterances, then analyzed them using a Grounded Theory approach~\cite{corbin2014grounded,muller2012grounded}. Two researchers independently coded each utterance as one of four ambiguity types (Referential, Temporal, Metric, Spatial), \textit{Out-of-frame} (outside the ambiguity taxonomy), or \textit{None} (no ambiguity). Of the 215 utterances, 207 were in agreement (96.3\%), yielding Cohen's $\kappa = 0.952$, indicating almost perfect inter-rater reliability~\cite{landis1977measurement}. The remaining 8 disagreements were resolved through discussion.
We reviewed and cleaned the transcribed utterances, then analyzed them using a Grounded Theory approach~\cite{corbin2014grounded,muller2012grounded}. 
\rev{The two researchers first open-coded the full set of utterances to identify what aspect of each query remained underspecified. During the process of comparing and merging similar codes, prior work on dialogue grounding~\cite{clark1991grounding} and ambiguity or underspecification in visualization NLIs~\cite{gao2015datatone,setlur2019inferencing} informed how we refined the codebook. The final codebook included category definitions and examples.}
The same two researchers independently coded each utterance as one of four ambiguity types (Referential, Temporal, Metric, Spatial), \textit{Out-of-frame} (outside the ambiguity taxonomy), or \textit{None} (no ambiguity). Of the 215 utterances, 207 were in agreement (96.3\%), yielding Cohen's $\kappa = 0.952$, indicating almost perfect inter-rater reliability~\cite{landis1977measurement}. The remaining 8 disagreements were resolved through discussion.

\subsection{Findings on Ambiguity in Spoken Queries}
\label{sec:findings}

Of the 215 utterances, 25 were classified as \textit{None} (no ambiguity) and 41 as \textit{Out-of-frame} (outside the ambiguity taxonomy, described below). Both were excluded from analysis. The remaining 149 utterances exhibited one of four ambiguity types. The most frequent was Metric (\(n=66\), 30.7\%), followed by Referential (\(n=56\), 26.0\%), Temporal (\(n=21\), 9.8\%), and Spatial (\(n=6\), 2.8\%). Rather than fully specifying who, when, where, or what metric they intended, participants produced short, context-dependent utterances whose interpretation depended on the current scene, prior discourse, and inferred viewing focus.

\textbf{Metric Ambiguity} (\(n=66\), 30.7\%). Metric ambiguity was the most prevalent type, occurring when participants implied a need for statistical or comparative information without specifying the exact metric, comparison target, or evaluation criterion. \textit{``How did the player's performance compare to Son?''} left open which metric (e.g., goals, assists, pass accuracy) should be compared. Users may have had a particular metric in mind but did not specify it, or may not have known which metrics were available. Other examples included \textit{``Please compare the stats of the attackers of each team''} and \textit{``Player 11 in black uniform, is he good?''}, where the scope and type of statistics were not explicit.

\textbf{Referential Ambiguity} (\(n=56\), 26.0\%). Referential ambiguity arose when the intended referent (e.g., a specific player, team, or event) was underspecified, often when multiple entities were simultaneously visible or recently mentioned. For example, \textit{``Who passed the ball to the player in the middle?''} becomes problematic when multiple passes occurred in quick succession. Similarly, \textit{``Who's the midfielder? The one in the black uniform''} or \textit{``What about the other one?''} could refer to multiple entities on the field, making it difficult for the system to determine the intended target from speech alone.

\textbf{Temporal Ambiguity} (\(n=21\), 9.8\%). Temporal ambiguity occurred when the intended time window or moment was unclear, particularly in replay-related requests. \textit{``Please replay the foul situation again''} is ambiguous both in which foul it refers to when multiple fouls occurred, and in where the replay should begin (the initial contact, the build-up, or the referee's decision). Similarly, \textit{``Summarize what just happened''} left ambiguous whether the user meant the last few seconds, the most recent play, or a longer sequence. This type was salient in sports contexts where multiple similar events occur throughout a match.

\textbf{Spatial Ambiguity} (\(n=6\), 2.8\%). Spatial ambiguity, though rare, arose when questions depended on spatial relations not fully specified in speech. \textit{``Why did the opponent foul in that area?''} leaves unclear which region ``that area'' refers to. \textit{``Isn't the position where the blue player received the ball offside?''} requires knowledge of the exact spatial layout at the moment of the pass. The low frequency suggests participants relied less on spatial descriptions, because spatial relations in sports are inherently dynamic and difficult to articulate verbally.

\textbf{Out-of-frame Utterances} (\(n=41\), 19.1\%). A notable portion of utterances fell outside our ambiguity taxonomy because they did not target factual, retrievable match information. These fell into three subcategories: (1)~\textit{tactical reasoning} (e.g., \textit{``What are the advantages of that formation?''}), (2)~\textit{rule inquiries} (e.g., \textit{``I wonder what the standard of yellow and red cards is in FIFA''}), and (3)~\textit{what-if or hypothetical} questions (e.g., \textit{``If you were the coach, what kind of feedback would you give?''}). While these represent legitimate user needs, they require subjective judgment or reasoning beyond observable match data. Their prevalence highlights that speech-driven sports-viewing systems must not only handle ambiguous factual queries but also gracefully manage requests extending into opinion, speculation, and domain reasoning.
 
Taken together, these findings suggest that ambiguity in speech-driven sports viewing often stems from underspecified reference, metric, time, and space rather than recognition errors. These four ambiguity types informed the design space presented in the next section.
 
% \textbf{User Feedback for Design.} We refer to formative study participants as FP1–FP5.Participants also provided feedback on what would help address these ambiguities. For referential queries, two participants (FP1, FP3) noted difficulty identifying the intended player from speech alone and suggested that displaying player names or candidate indicators directly on or above players in the 3D scene would help resolve who was being referred to. For temporal queries, FP4 noted that specifying a moment was difficult when events occurred in close succession, and suggested that seeing an event sequence would make the intended moment identifiable. For metric queries, a notable observation was that participants were often unaware of their own scope ambiguity: FP2 noted that asking about statistics without specifying the time frame (e.g., this match vs. this season) produced answers that felt incorrect, even though the system had made a plausible assumption. These observations informed our externalization designs: in-situ player badges for referential cases, timeline thumbnail sequences for temporal cases, and explicit scope panels for metric cases, each aimed at making the system's assumed interpretation inspectable at the point where ambiguity arose.

\textbf{User Feedback for Design.} Formative study participants (FP1--FP5) provided feedback on addressing these ambiguities. For referential queries, FP1 and FP3 suggested displaying player names or candidate indicators directly on players in the 3D scene to resolve who was being referred to. For temporal queries, FP4 noted that specifying a moment was difficult when events occurred in close succession, and suggested that seeing an event sequence would help. For metric queries, FP2 observed that asking about statistics without specifying the time frame (e.g., this match vs. this season) produced answers that felt incorrect, even though the system had made a plausible assumption, revealing that participants were often unaware of their own scope ambiguity. These observations informed our externalization designs: in-situ player badges for referential cases, timeline thumbnail sequences for temporal cases, and explicit scope panels for metric cases.
\section{Design Space}

% One thing: what kind of question supporting! - not tactic one or other reasoning of other ppl thoughts? Ambiguity of reference in context

% \\
% None - ambiguity / types of answers for reasoning
% Replacing - reasoning - out of frame
% \\

\begin{figure}[t]
    \centering
    \includegraphics[width=\linewidth]{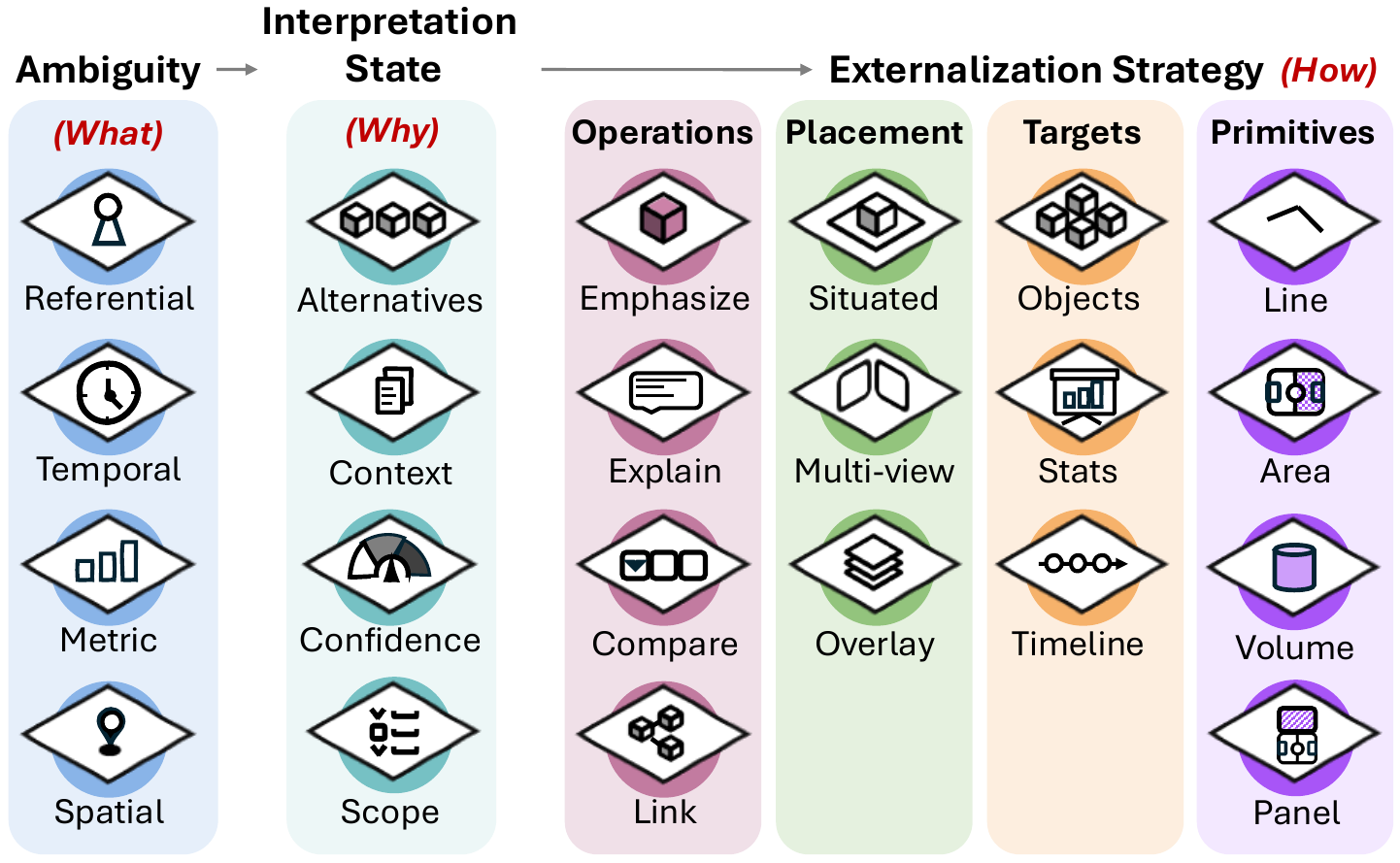}
    \caption{Design space of externalizing ambiguous speech in XR sports viewing. The figure organizes the design space into ambiguity types, interpretation state, and externalization strategy.}
    \label{fig:design_space}
    \vspace{-1em}
\end{figure}
 
\begin{figure*}[t]
    \centering
    \includegraphics[width=1.0\linewidth]{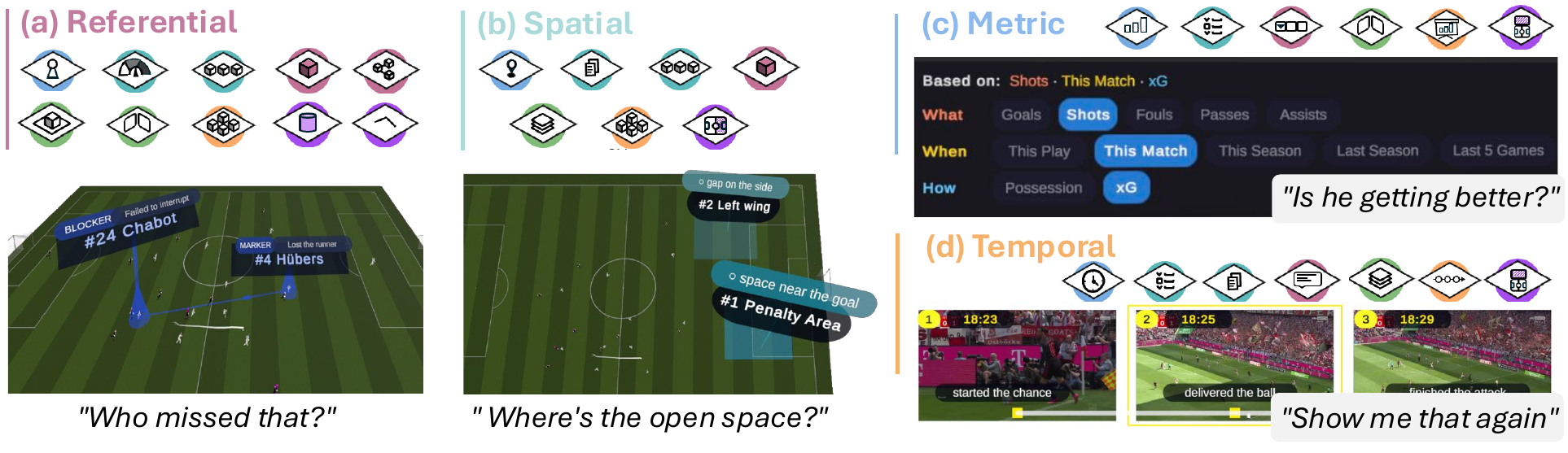}
    \caption{Four representative externalization cases: (a) referential with candidate badges, (b) spatial with zone overlays, (c) metric with scope panel, and (d) temporal with timeline thumbnails. Each illustrates how the system surfaces its interpretation within the viewing environment.}
   
    \label{fig:cases}
    \vspace{-1em}
\end{figure*}

Our design space focuses on factual, match-relevant queries (i.e., those classified as Referential, Temporal, Metric, or Spatial in our formative study), while Out-of-frame utterances and non-ambiguous queries fall outside the current scope. Building on the ambiguity types identified in our formative study (Sec~\ref{sec:findings}), our design contribution lies less in defining these types themselves, which parallel established categories in dialogue systems and NLP~\cite{clark1991grounding, min2020ambigqa, setlur2019inferencing}, and more in identifying what aspects of a system's committed interpretation can be externalized. Prior work on speech disambiguation has largely focused on resolving ambiguity before the system acts. Our framework instead addresses what to reveal after the system commits to one reading, shifting the design challenge from preventing ambiguity to making post-commitment interpretations transparent. 
% We organize our design space along three dimensions: the \textbf{ambiguity type}, which characterizes \textbf{\textit{what}} remains underspecified in the user's query, the \textbf{interpretation state}, which identifies \textbf{\textit{why}} the system's interpretation warrants externalization, and the \textbf{externalization strategy}, which specifies \textbf{\textbf{how}} to present it within the XR environment (Fig.~\ref{fig:design_space}).
We organize our design space along three dimensions: the \textbf{ambiguity type}, which characterizes \textbf{\textit{what}} remains underspecified in the user's query, the \textbf{interpretation state}, which identifies \textbf{\textit{why}} the system's interpretation warrants externalization, and the \textbf{externalization strategy}, \rev{which draws on situated and XR visualization design~\cite{lo2021xrspectator, xu2023xair} to specify \textbf{\textit{how}} to present it within the XR environment (Fig.~\ref{fig:design_space}).}

\subsection{Ambiguity Types}
Our design space builds on the four ambiguity types identified in the formative study (Section~\ref{sec:findings}): \textit{Referential} (which entity?), \textit{Temporal} (which moment?), \textit{Metric} (which statistic?), and \textit{Spatial} (which region?). Each type describes a distinct dimension that may remain underspecified in a user's spoken query, \rev{aligning with prior work on dialogue grounding, ambiguous questions, and underspecified natural-language interaction~\cite{clark1991grounding, min2020ambigqa, setlur2019inferencing}.} In the following, we describe how our system detects and resolves each type.

\subsection{Interpretation State to Externalize}
When a system interprets an ambiguous spoken query, several conditions make that interpretation worth surfacing. 
% Rather than treating interpretation as a hidden step that produces a final answer, our framework treats it as something that can be externalized through embedded visualizations in XR: 
\rev{Following prior NLI work on exposing ambiguity and system decisions for user correction~\cite{gao2015datatone, shen2024xnli}, our framework treats interpretation as something that can be externalized through embedded visualizations in XR rather than as a hidden step that produces a final answer:}
the system may have considered other candidates, relied on contextual assumptions, committed with varying confidence, or assumed a  scope. Each represents a reason why the interpretation warrants externalization. We identify four corresponding forms of interpretation state: alternatives, context, confidence, and scope.

\textbf{Alternatives.}
A system may identify multiple plausible interpretations of the same query, such as several candidate referents. Externalizing alternatives can help users inspect what the system considered possible rather than only seeing the final interpretation. For example, when a user asks \textit{``Who's the midfielder?''} and multiple midfielders are on the field, the system may surface the top candidates it considered, allowing the user to select the intended one. Even when the system's top-ranked interpretation is incorrect, presenting alternatives gives users a path to recover the intended answer without reformulating their query.

\textbf{Context.}
Context represents the basis on which the system selects its interpretation. Interpretation often depends on contextual assumptions, including the user's previous questions in the conversation, the current game state tracked by the system (e.g., player positions, ball location, ongoing events), or the active replay segment. Externalizing context can help users understand why a particular interpretation was selected. In spatial clarification, for example, the system may interpret \textit{``Why did they foul in that area?''} based on tracked player and ball positions at the time of the foul, and surface which region it inferred so the user can confirm or redirect. In moment retrieval, the system may interpret \textit{``Please replay the foul''} as referring to the most recent foul based on recency in the event timeline. Surfacing these contextual assumptions allows users to understand the reasoning behind the system's selection and correct it when the wrong event or entity was chosen.

\textbf{Confidence.}
Even when a system selects one interpretation, it may have varying confidence about whether that interpretation matches the user's intent. Externalizing confidence is particularly relevant in identity resolution, where the system must determine which entity the user is referring to among multiple candidates. For instance, the system might indicate high confidence when only one player matches the query, but lower confidence when several candidates are plausible. Making confidence visible, the system signals whether its interpretation is reliable, inviting the user to confirm or correct if needed.

\textbf{Scope.}
While context determines \textit{which} interpretation the system selected, scope determines \textit{how much} of the response is presented. A query may be ambiguous in both the unit of analysis and the range of data considered. In performance comparison, scope operates along two dimensions: who is being analyzed (e.g., an individual player, a positional group, or an entire team) and what range of data is considered (e.g., this match, the current season, or career statistics). For example, \textit{``How is he doing?''} leaves both dimensions open, and the system may surface its assumed unit and time range so the user can adjust. In moment retrieval, scope determines the temporal extent of the replay, such as whether to show only the key action or include the surrounding build-up and aftermath. Surfacing the assumed scope allows users to expand or narrow the system's framing to match their intent.

\subsection{Externalization Strategy}
\rev{Given the ambiguity type and interpretation state, an externalization strategy determines how the response is presented in XR, drawing on situated visualization and XR explanation design~\cite{willett2016embedded, xu2023xair}.} The design space covers four sub-dimensions: operations, placement, targets, and primitives (Fig.~\ref{fig:design_space}).

\textbf{Operations.}
Operations describe the communicative function of the externalization. We identify four operations: \textit{Emphasize} draws the user's attention to a specific entity or region, such as highlighting the player the system believes the user is referring to. \textit{Explain} provides a structured account of events surrounding the selected interpretation, such as overlaying a timeline showing the sequence around the selected moment. \textit{Compare} presents multiple data points side by side, such as showing different metrics or scopes for the user to evaluate. \textit{Link} establishes visual connections between related entities, such as connecting a player to their associated event.

\textbf{Placement.}
Placement describes where in the XR environment the externalization is rendered. We identify three placement strategies: \textit{Situated} places visualizations directly on or near an entity in the scene, such as attaching an information panel above a player that follows their position. \textit{Multi-view} places information in a separate, fixed region of the XR environment, such as a side panel or tabletop display that persists independently of the video content. \textit{Overlay} renders information spanning a broader region, such as displaying event markers along a timeline or highlighting a zone on the field.

\textbf{Targets.}
Targets describe what type of data the externalization acts upon. We identify three target types: \textit{Objects} refers to entities in the scene such as players, teams, or the ball, typically the focus of referential and spatial queries. \textit{Stats} refers to quantitative data such as pass accuracy or shot counts, typically the focus of metric queries. \textit{Timeline} refers to temporal sequences or event histories within the match, typically the focus of temporal queries.

\textbf{Primitives.}
Primitives describe the visual elements used to render the externalization in XR. Building on the mark taxonomy~\cite{munzner2014visualization}, which classifies visual marks by spatial dimensionality, we identify four primitives adapted for XR environments: \textit{Line} (1D) uses linear marks such as arrows or trajectories to indicate direction, movement, or relationships. \textit{Area} (2D) uses two-dimensional regions such as highlights or shaded surfaces to indicate spatial extent or emphasis. \textit{Volume} (3D) uses three-dimensional shapes or bounded regions to enclose or emphasize objects. \textit{Panel} uses flat information displays such as stat cards or comparison tables, serving as a compound container for text and data.

\subsection{Mapping Ambiguity Types to Externalization Design}
% To illustrate how these dimensions combine in practice, we
% instantiate four representative cases, each grounded in one
% ambiguity type (Figure~\ref{fig:cases}). Section~\ref{sec:visual_externalization}
% details how each case maps design space dimensions to
% implemented visualizations and describes their iterative
% refinement (Figure~\ref{fig:iterations}).
\rev{To clarify the scope of the design space, we present it as an organizing framework for representative instantiations, rather than as an exhaustive evaluation of all possible combinations. Figure~\ref{fig:cases} shows four representative and non-exhaustive implemented cases, each grounded in one ambiguity type. The icons above each case indicate the selected design-space dimensions across ambiguity type, interpretation state, and externalization strategy.}
Section~\ref{sec:visual_externalization}
describes the implementation of these cases and their iterative
refinement (Figure~\ref{fig:iterations}).
\section{System}

\begin{figure}[t]
    \centering
    \includegraphics[width=\linewidth]{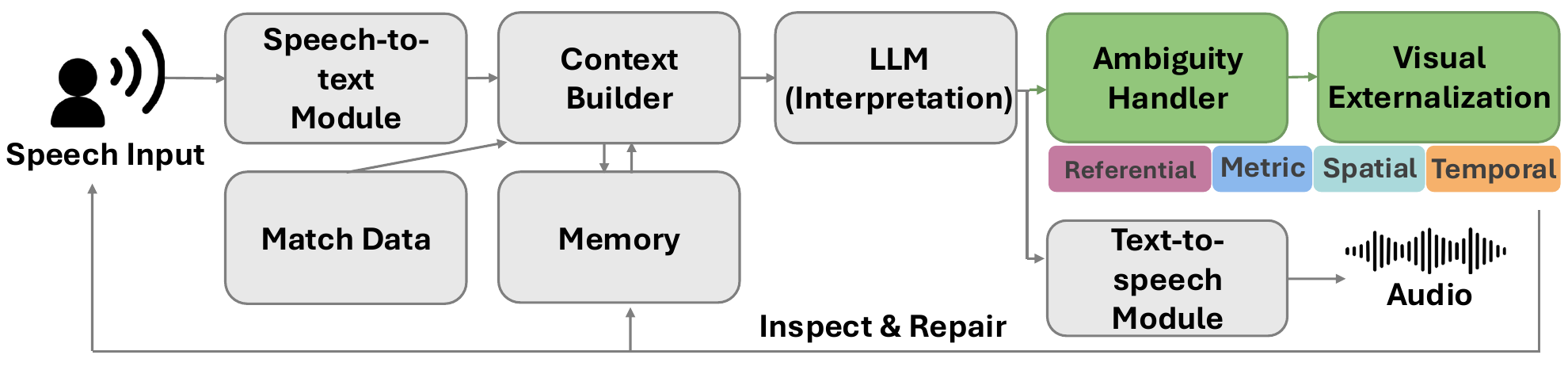}
    \caption{System pipeline for speech-driven XR sports viewing with ambiguity-aware interpretation and visual externalization.}
    \label{fig:system}
    \vspace{-1em}
\end{figure}

To examine how system interpretations of ambiguous speech can be externalized in XR sports viewing, we developed an XR soccer viewing system that combines speech-based querying, ambiguity-aware interpretation, and coordinated visual externalization across immersive and supporting views (\cref{fig:system}). Rather than returning an answer, the system externalizes how it interpreted the query, allowing users to inspect assumptions and correct mismatches through follow-up queries.

\subsection{Data and Scene Construction}
Following prior work that uses pre-recorded match data for interactive sports viewing research~\cite{lin2022omnioculars, lee2024sportify, chen2023iball}, our system is built on the Integrated Dataset of Spatiotemporal and Event Data in Elite Soccer (IDSSE)~\cite{bassek2025integrated}, a publicly available dataset covering seven German Bundesliga matches. It provides x/y trajectories for all players and the ball at 25\,Hz, synchronized event annotations (passes, shots, fouls, set pieces), and match-level metadata (Table~\ref{tab:data}). We selected two matches (1.~FC K{\"o}ln vs.\ FC Bayern M{\"u}nchen and VfL Bochum~1848 vs.\ Bayer~04 Leverkusen) for the study. The XR scene reconstructs player and ball movement from trajectory data, while supporting views provide coordinated analytic context such as event-aligned timelines and summary statistics.

\vspace{-0.5em}
\begin{table}[t]
\centering
\caption{Data types from the IDSSE dataset~\cite{bassek2025integrated}.}
\label{tab:data}
\small
\begin{tabular}{@{}ll@{}}
\toprule
\textbf{Type} & \textbf{Description} \\
\midrule
Position Data     & x/y-coordinates of all players and the ball at 25\,Hz \\
Event Data        & Timestamped events (passes, shots, fouls) with player and outcome \\
Match Metadata    & Teams, players, jersey numbers, formations \\
Ball State        & Per-frame possession (home/away) and play status (active/inactive) \\
\bottomrule
\end{tabular}
\vspace{-1em}
\end{table}

\subsection{Speech Query Processing}
The system accepts spoken queries during match viewing via push-to-talk: pressing and holding a controller button initiates recording, and releasing it sends the audio to OpenAI Whisper for transcription. We chose push-to-talk over continuous listening for clear utterance segmentation and reduced latency compared to server-side voice activity detection. The transcribed query is passed to a context builder, which assembles the current match state from trajectory and event data (Table~\ref{tab:data}): player positions at the current frame, recent events, and player metadata. The context builder incorporates a conversation memory that retains recent query-response pairs (adjustable to multiple turns), enabling resolution of follow-up references. The resolved interpretation from each turn, including identified player IDs, event references, selected metrics, and time windows, is injected into the subsequent LLM prompt as conversation context, supporting pronominal references (e.g., ``What about his assists?'') and references to prior events or metric selections (e.g., ``Was that a foul?'' or ``Compare their xG instead?'').

\begin{figure*}[t]
    \centering
    \includegraphics[width=\linewidth]{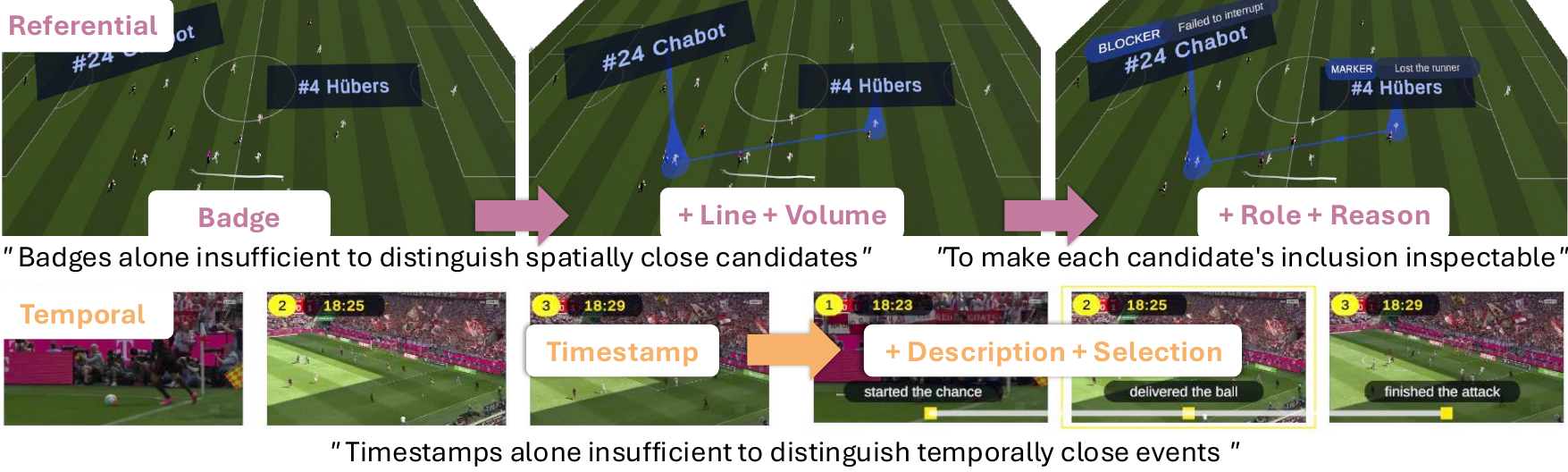}
    \caption{Design iterations for referential and temporal externalization. Each row shows progressive refinements addressing pilot feedback, from minimal cues (badge or timestamp only) to the final design with added visual and textual context.}
    \label{fig:iterations}
    \vspace{-1.5em}
\end{figure*} 

\subsection{Visual Externalization Design}
\label{sec:visual_externalization}

When the ambiguity handler classifies a query into one of
the four ambiguity types, the system generates a corresponding
visual externalization. Each design was developed through
iterative refinement incorporating feedback from the formative
study and pilot sessions (Figure~\ref{fig:iterations}).

\textbf{Identity Resolution (Referential).}
The system externalizes its \textit{confidence} in the selected referent and surfaces \textit{alternatives} when multiple candidates are plausible. The primary operations are \textit{emphasize} and \textit{link}, rendered as \textit{situated} visualizations on candidate players using \textit{line} and \textit{volume} primitives. Each candidate is highlighted with a badge displaying jersey number and name, with the top-ranked candidate positioned highest to encode the system's relative confidence. In early iterations, badges alone were insufficient when multiple players were spatially close. We addressed this by adding directional lines connecting candidates to encode relationships (e.g., a pass trajectory) and semi-transparent volume primitives (cone beams) to distinguish candidates from surrounding players. We further added a role label in the interpreted event (e.g., \textsc{blocker}, \textsc{marker}) and a brief reason alongside each badge (e.g., ``failed to intercept'') to make each candidate's inclusion inspectable.  Badges are billboarded to face the user's camera and staggered in height to reduce occlusion.

\textbf{Spatial Clarification (Spatial).}
The system externalizes \textit{context} and \textit{alternatives} through \textit{overlay} placement using \textit{area} primitives targeting \textit{objects}. Labeled zone overlays are rendered onto the field surface, each demarcated with a semi-transparent area annotated with a zone label (e.g., ``Left wing,'' ``Penalty Area'') and a ranking number, with higher-ranked zones at greater opacity. Initially, overlays displayed only zone names, but pilot sessions showed this was insufficient. We added a brief contextual reason to each label (e.g., ``gap on the side'') to communicate the basis of the system's spatial interpretation.

\textbf{Performance Comparison (Metric).}
The system externalizes the assumed \textit{scope} through \textit{compare} operations placed in \textit{multi-view} panels using \textit{panel} primitives targeting \textit{stats}. The scope is shown along three dimensions: \textit{What} metric (e.g., goals, shots, passes), \textit{When} the data is drawn from (e.g., this match, this season), and \textit{How} it is measured (e.g., possession, xG). Unlike conventional option menus, the panel displays the system's current interpretation with pre-selected values highlighted, making explicit which assumptions the system made. A header line summarizes the active interpretation (e.g., ``Based on: Shots $\cdot$ This Match'') so users can verify the framing at a glance and adjust any dimension through follow-up queries.

\textbf{Moment Retrieval (Temporal).}
The system externalizes \textit{scope} and \textit{context} through \textit{explain} operations rendered as \textit{overlays} targeting the \textit{timeline} using \textit{panel} primitives. A horizontal sequence of candidate replay segments is presented, each shown as a thumbnail annotated with a timestamp. Segments are ordered chronologically and numbered, with the system's selected segment emphasized via a highlighted border. In early iterations, timestamps alone were insufficient to distinguish temporally close events. We addressed this by adding a contextual description to each thumbnail (e.g., ``started the chance,'' ``delivered the ball'') to clarify what each segment depicts.

\subsection{Implementation and Computational Evaluation}
\label{sec:implementation}

The system was developed in Unity 6000.2.6f2 and deployed on Meta Quest 3. Speech-to-text uses OpenAI Whisper via API, and ambiguity classification and interpretation are handled by GPT-4o, which returns a structured JSON response parsed by the ambiguity handler. 
\rev{To reduce unsupported outputs, we used deterministic decoding, constrained responses to a structured JSON schema, grounded prompts in retrieved match data, and parsed only supported fields for visualization.}
% \rev{We used deterministic decoding, JSON-schema responses, match-data grounding, and supported-field parsing to reduce unsupported or variable outputs.}
(The full system prompts are provided in the supplementary material.) The end-to-end response latency averaged $M = 2.77$s ($SD = 0.97$s, median $= 2.58$s) across all non-tutorial trials.

To assess classification reliability, we used the 215 formative study utterances with consensus labels from two independent coders as ground truth, merging \textit{None} and \textit{Out-of-frame} into a single \textit{None} category. We ran each utterance through GPT-4o with the same system prompt to compare against these labels. The system achieved 86.5\% accuracy (Cohen's $\kappa = 0.815$), with per-class F1-scores ranging from 0.800 (Spatial) to 0.952 (Temporal). The primary source of misclassification was \textit{None} $\rightarrow$ \textit{Metric} (12 cases), where out-of-frame queries containing statistical language were classified as metric ambiguity (Table~\ref{tab:confusion_matrix}). This is consistent with our formative finding that tactical and hypothetical questions often include performance-related vocabulary that overlaps with genuine metric queries. 
\rev{Misclassifications among the four ambiguity types were infrequent, but when they occur, the system may render the wrong externalization template, such as badges instead of a timeline.}
% Misclassifications among the four ambiguity types themselves were infrequent, suggesting reasonable reliability for interactive use.
 
% \begin{table}[t]
% \centering
% \caption{Confusion matrix for ambiguity classification (GPT-4o, $N=215$). Out-of-frame queries are mapped to \textit{None}. Rows represent ground-truth labels; columns represent predicted labels.}
% \label{tab:confusion_matrix}
% \begin{tabular}{l|ccccc|c}
% \toprule
%  & \textbf{Ref.} & \textbf{Temp.} & \textbf{Metr.} & \textbf{Spat.} & \textbf{None} & \textbf{Total} \\
% \midrule
% \textbf{Referential}  & \textbf{49} & 0  & 4  & 0 & 3  & 56 \\
% \textbf{Temporal}      & 1  & \textbf{20} & 0  & 0 & 0  & 21 \\
% \textbf{Metric}   & 2  & 1  & \textbf{63} & 0 & 0  & 66 \\
% \textbf{Spatial}       & 0  & 0  & 0  & \textbf{4} & 2  & 6 \\
% \textbf{None}          & 4  & 0  & 12 & 0 & \textbf{50} & 66 \\
% \bottomrule
% \end{tabular}
% \vspace{-1em}
% \end{table}

\begin{table}[t]
\centering
\caption{Confusion matrix for ambiguity classification (GPT-4o, $N=215$). Out-of-frame queries are mapped to \textit{None}. Rows represent ground-truth labels; columns represent predicted labels.}
\label{tab:confusion_matrix}
\vspace{-0.6em}
\scriptsize
\setlength{\tabcolsep}{3pt}
\renewcommand{\arraystretch}{0.85}
\begin{tabular}{@{}l|ccccc|c@{}}
\toprule
 & \textbf{Ref.} & \textbf{Temp.} & \textbf{Metr.} & \textbf{Spat.} & \textbf{None} & \textbf{Tot.} \\
\midrule
\textbf{Ref.}  & \textbf{49} & 0  & 4  & 0 & 3  & 56 \\
\textbf{Temp.} & 1  & \textbf{20} & 0  & 0 & 0  & 21 \\
\textbf{Metr.} & 2  & 1  & \textbf{63} & 0 & 0  & 66 \\
\textbf{Spat.} & 0  & 0  & 0  & \textbf{4} & 2  & 6 \\
\textbf{None}  & 4  & 0  & 12 & 0 & \textbf{50} & 66 \\
\bottomrule
\end{tabular}
\vspace{-1em}
\end{table}
\section{User Study}
 
We conducted a within-subjects study ($N=16$) comparing a baseline condition (voice-only) against an externalization condition (voice response with situated visual cues). 

\subsection{Study Design and Procedure}

\textbf{Design.} Each participant completed both conditions in counterbalanced order (2$\times$2 Latin square: condition order $\times$ video set). Each condition contained 4 video clips from two Bundesliga matches, with 2 sequential scenarios per clip (8 total per condition). Scenarios varied in whether the system's interpretation was likely to align or diverge from the participant's intent, based on match context (e.g., number of nearby players, recent events). Video assignments were counterbalanced across conditions to control for content effects. 

\textbf{Task.} Participants watched short soccer clips (15--30s, looping) on a tabletop 3D pitch reconstruction and asked questions using voice. For each clip, a brief scenario description oriented participants toward match-relevant queries (e.g., ``Bayern just scored through a passing move. You might want to ask about the attack.''). Participants were briefed that the system may not always interpret correctly and were instructed to evaluate responses against the video and attempt corrections through follow-up queries.

% \textbf{Conditions.} In the externalization condition, the system's response was accompanied by situated visual cues corresponding to the detected ambiguity type: player candidate badges with role labels (referential), zone overlays on the field surface (spatial), timeline thumbnail sequences (temporal), or scope selection panels (metric). In the baseline condition, participants received only the voice response.

\label{sec:conditions}
\textbf{Conditions.} In the externalization condition, the system's response was accompanied by situated visual cues corresponding to the detected ambiguity type: player candidate badges with role labels (referential), zone overlays on the field surface (spatial), timeline thumbnail sequences (temporal), or scope selection panels (metric). In the baseline condition, participants received only the voice response. We chose voice-only as the baseline because it reflects the current state of speech-based XR sports viewing, where no surveyed application exposes interpretation processes (Section~\ref{sec:apps}), and because comparing explanation-present against explanation-absent conditions is the recommended baseline design in XAI user studies~\cite{rong2024human}, consistent with prior work on transparency in human-AI decision-making~\cite{bucinca2021trust, lai2019human}.

\textbf{Aligned and Misaligned Trials.} Each scenario was pre-designated as either aligned (system interpretation matches the participant's likely intent) or misaligned (steered toward a plausible but non-primary interpretation). We included both types because relying solely on naturally occurring misinterpretations would have produced sparse and uneven trial counts across ambiguity types, precluding per-type exploratory analyses. Following prior work that deliberately manipulates system outputs to study detection and repair behavior~\cite{lakkaraju2020fool, heuer2020accuracy}, misalignment was induced by appending a scenario-specific constraint to the system prompt while leaving the pipeline otherwise unchanged. For instance, in a referential scenario where Bayern scored through a passing sequence, the aligned condition resolved ``Who passed it?'' to the most salient passer, while the misaligned condition selected a different but plausible passer from earlier in the sequence. For a temporal replay query, the misaligned condition shifted the starting point so that the key moment fell outside the initial view. In all cases, misaligned interpretations remained within the space of plausible readings. Aligned trials served as an ecological validity check for whether externalization promotes inspection beyond error correction alone.
\rev{Out-of-scope queries were mapped to \textit{None} and excluded from aligned/misaligned analyses.}

\textbf{Procedure.} Sessions lasted approximately 60 minutes. Participants first completed a tutorial in which they practiced asking questions and receiving system responses. In the main phase, they completed both conditions in counterbalanced order. After both conditions, participants watched a longer match segment (approximately 5 minutes) and freely explored the system without guided scenarios, followed by a post-session questionnaire and semi-structured interview.

\subsection{Participants}
 
We recruited 16 participants (11 male, 5 female; aged 22--40, $M = 29.6$, $SD = 4.7$) with diverse soccer backgrounds: 8 watched rarely, 4 occasionally, and 4 frequently.
This range reflects viewers who seek match information but may lack expertise to construct precise queries, the target population for speech-driven viewing systems.
Soccer familiarity ranged from not familiar (1) to very familiar (3), with most reporting slight (7) or moderate (5) familiarity. Twelve (75\%) had previously looked up stats during games. XR experience ranged from never (5) to regular use (2), with the remainder having tried a headset once or twice (5) or a few times (4). Most participants used voice assistants rarely (10) or sometimes (3), with 2 reporting never and 1 often.
The study protocol was reviewed and approved by the institutional review board at Harvard University (Protocol No. IRB25-0906). All participants provided informed
consent prior to participation.
 
\subsection{Measures}
\label{sec:measures}

We collected behavioral, subjective, and qualitative measures.

\textbf{Quantitative.} From interaction logs, we extracted: (1)~\textit{repair utterance rate}, the proportion of queries containing correction or redirection language, identified via a keyword-based classifier with 13 word-boundary patterns across three categories: \textit{explicit correction} (``i mean,'' ``not that,'' ``actually''), \textit{negation and redirection} (``wrong,'' ``instead,'' ``other one,'' ``not the''), and \textit{re-request} (``specific,'' ``correct,'' ``you said''), 
and (2)~\textit{confirmation time}, the interval from the system's first response to the participant's next query. \rev{Our main question was whether externalization encourages repair without additional costs.}

\textbf{Subjective and Qualitative.} Participants rated each condition on a 7-point Likert scale across four inspectability items (ease of identifying which player the system referred to, which moment it showed, what went wrong, and what alternatives it considered), along with items for trust, perceived clutter, and overall satisfaction. Inspectability items were informed by XAI transparency frameworks~\cite{mohseni2021multidisciplinary, hoffman2023measures}. Trust was adapted from~\cite{jian2000foundations, scharowski2025trust} and ambiguity awareness from~\cite{clark1991grounding}. Semi-structured interviews (10 min) covered misunderstanding moments, repair strategies, use of externalized cues, and visual clutter trade-offs, analyzed via thematic analysis~\cite{braun2006thematic}.
% no dif - rigruous / starting with p-value
% no p-value is later - need to show obvious one
% the most obvious a, b, c / no interaction cost / weakness of our system 
% some how link visualization results - hard to interpret - somehow make people easily understand - benefitial for finding my findings. 
% rather than showing bbox, need to make confidence score + bar chart
% quantative (rate, costs) - bar chart
% claim -> numbers -> discussion -> other papers -> comments
% design implication parts 

\begin{figure*}[t]
\centering
\includegraphics[width=1.0\linewidth]{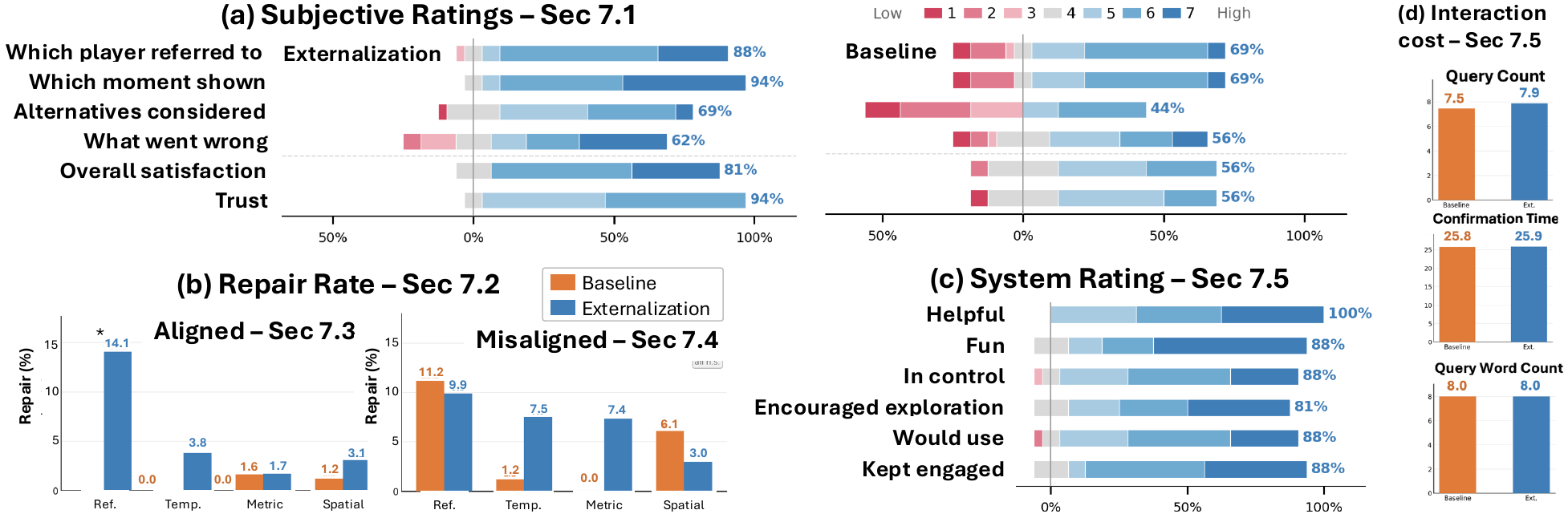}
\caption{Study results. \textbf{(a) Subjective Ratings}: Diverging Likert distributions for externalization \rev{(left)} and baseline \rev{(right)}. $*$ $p < .05$ (Wilcoxon). \rev{(b) Repair Rate: Proportion of repair-language queries by ambiguity type in aligned (left) and misaligned (right) trials.} \textbf{(c) System-Level Ratings}: Post-session usability ratings (externalization condition). \rev{ \textbf{(d) Interaction Cost}: query count, confirmation time, and query word count.} }
\label{fig:results}
\vspace{-1.5em}
\end{figure*}

\section{Results}
We report findings from 16 participants across 256 trial scenarios ($16 \times 8$ video clips $\times$ 2 scenarios per clip). All pairwise comparisons used Wilcoxon signed-rank tests on per-participant paired means, and effect sizes are reported as $r = |Z|/\sqrt{N}$.

\subsection{Externalization Helps Users Understand What the System Assumed} The intended benefit of externalizing system interpretations is that users can inspect what the system decided, not just what it returned. We measured this through four Likert inspectability items adapted from XAI transparency frameworks~\cite{hoffman2023measures, mohseni2021multidisciplinary}. Inspectability was significantly higher under externalization on three of four items (Figure~\ref{fig:results}a): identifying which player the system referred to ($M_{\text{ext}} = 5.94$ vs.\ $M_{\text{base}} = 4.69$; $W = 3.0$, $p = .011$, $r = .62$), which moment was shown ($M_{\text{ext}} = 6.25$ vs.\ $M_{\text{base}} = 4.69$; $W = 4.0$, $p = .003$, $r = .73$), and what alternatives were considered ($M_{\text{ext}} = 4.75$ vs.\ $M_{\text{base}} = 3.62$; $W = 5.5$, $p = .024$, $r = .56$). Identifying what went wrong did not reach significance ($M_{\text{ext}} = 5.06$ vs.\ $M_{\text{base}} = 4.31$; $W = 10.5$, $p = .151$, $r = .36$), suggesting that externalization may support recognizing what the system assumed more readily than diagnosing why it erred, though this difference requires further investigation. Overall satisfaction was significantly higher under externalization ($M_{\text{ext}} = 5.69$ vs.\ $M_{\text{base}} = 4.56$; $W = 0.0$, $p = .002$, $r = .73$), while trust and perceived clutter did not differ between conditions (Appendix~\ref{appendix:subjective}), indicating that the benefit was perceptual and experiential rather than a generalized increase in system confidence or an attentional burden. P02 noted that externalization made it \textit{``easier to see how the system understood my question or what it chose to show me,''} leveraging externalized cues to guide subsequent queries rather than passively accepting responses.

\subsection{Repair Language is Higher Under Externalization} 
% \rev{Given that externalization support higher inspectability, we next examine whether it were also associated with repair of misaligned interpretations.}
\rev{We next examine whether these inspectability gains led to repairs under misaligned interpretations.}
We measured repair behavior as the proportion of queries containing explicit repair language, identified via the keyword-based classifier described in Section ~\ref{sec:measures}. Because this classifier does not detect implicit strategies such as query narrowing or rephrasing, the reported rates represent a conservative lower bound. Repair utterance rate was significantly higher under externalization ($M_{\text{ext}} = 5.6\%$, $SD = 3.8\%$) than baseline ($M_{\text{base}} = 3.3\%$, $SD = 4.2\%$; $W = 16.0$, $p = .039$, $r = .57$) (Figure~\ref{fig:results}b). As a manipulation check, participants produced more follow-up queries in misaligned trials ($M_{\text{mis}} = 33.75$) than aligned trials ($M_{\text{align}} = 27.62$; $W = 13.0$, $p = .004$), confirming that misalignment produced observable behavioral differences and suggesting that the increase reflected targeted correction rather than undirected query generation. P01 recalled: \textit{``I first asked `what team has better stats?'} and later asked \textit{`what team has better stats in the second half of the game?''' P05 reported ``learning over time to ask players' numbers and names''} to reduce ambiguity.
 
\subsection{Externalization Promotes Verification in Aligned Trials} To explore how repair behavior varied across ambiguity types, we conducted per-type comparisons. Given the small per-cell sample sizes from crossing ambiguity type with alignment condition, these analyses are exploratory and should be interpreted as indicative patterns rather than confirmatory findings. In particular, effect sizes may be informative where $p$-values lack power, and we report both throughout to support calibrated interpretation. The overall increase in repair utterance rate masks an important asymmetry: repair behavior differed between aligned and misaligned trials and across ambiguity types.

In aligned trials, repair utterance rate under externalization was significantly higher than baseline for referential queries ($M = 0.0\%$ vs.\ $14.1\%$; $W = 0.0$, $p = .018$, $r = .89$). This increase likely reflects that externalization provided a visible commitment against which participants could check their own intent. Under baseline, the system's referent selection was not surfaced, leaving no anchor for verification or challenge. While we cannot fully separate this from a visual-richness account, the direction of the effect (increased repair rather than increased confirmation) is consistent with grounding theory~\cite{clark1991grounding} and suggests that externalization prompted active inspection rather than passive consumption. No significant difference was observed for the remaining three types (all $p > .05$; Appendix~\ref{appendix:repair_aligned}), though the numerical trend under externalization was equal or higher than baseline across all four types. Under externalization, participants referenced visible interpretation elements directly:

\begin{quote} \vspace{-0.75em}  \textit{Externalization [P09, referential]:}\\ USER: ``How often does he score a goal?''\\ SYS: ``Jeremie Frimpong has scored 9 goals this season.''\\ \textit{[badges: Frimpong, Förster shown]}\\ USER: ``Not that player, but the player from the blue team.''\\ SYS: ``Philipp Förster scored 3 goals for the away team.'' \vspace{-0.75em}  \end{quote}

Twelve participants (75\%) found candidate displays useful for verifying referential interpretations.
\rev{P12: \textit{``it was really hard to follow which player the system was referring to due to similar uniforms.''}}

\subsection{Visibility Does Not Guarantee Repair: The Role of Correction Pathways}
Although externalization improved inspectability, it did not reliably increase correction in misaligned trials, where the system interpretation was shifted to a plausible but non-primary reading. No ambiguity type showed a significant repair increase (Appendix~\ref{appendix:repair_misaligned}). Referential and spatial repair remained similar across conditions (both $p > .5$). Temporal and metric queries showed large but nonsignificant effects (temporal: $r = .64$, $p = .194$; metric: $r = .91$, $p = .066$), suggesting underpowered trends that require replication. Repair occurred in only $38\%$ of misaligned externalization scenarios, leaving a $62\%$ visibility--action gap, consistent with XAI findings that visibility alone does not guarantee correction~\cite{bucinca2021trust, nourani2021anchoring}. \rev{Visual information alone may not explain the results. However, because the baseline was voice-only, we cannot separate visual richness from interpretation externalization.} Interpretation visibility may support inspection, but correction depends on repair cost when users must provide precise verbal specifications.

The visibility--action gap varied by ambiguity type, suggesting that repair cost was a key constraint. Temporal and metric queries showed higher-repair trends, but requiring precise verbal reformulation rather than direct selection may have limited correction, as P04 illustrates:

\begin{quote} \vspace{-0.75em} \textit{Externalization [P04, temporal]:}
\\
USER: `When did that, uh\ldots''\\ SYS: `Y. Sommer passed to M. de Ligt at 18:39.'' \\\textit{[temporal timeline shown]} \\ USER: ``When did Bayern score their first goal?'' \vspace{-0.75em} \end{quote}

For spatial queries, zone overlays may have appeared close enough to accept  ($M = 3.0\%$ vs.\ baseline $6.1\%$, $p = .686$), although the small number of trials warrants caution. Referential candidate badges likewise did not increase repair ($M = 9.9\%$ vs.\ $11.2\%$, $p = .611$), consistent with anchoring to visible system outputs~\cite{nourani2021anchoring}. Thus, externalization may shape repair tendencies, but effective correction requires lower-cost, ambiguity-specific interactions, such as tapping candidate badges, scrubbing timelines, or toggling metric scopes~\cite{lee2021whatsthis, lee2024gazepointar}.

\subsection{Externalization Imposes No Additional Interaction Cost}
A practical concern is whether these benefits come at the cost of additional interaction overhead. We compared query count per scenario, response-to-next-query interval, and query length between conditions. None differed significantly (all $p > .05$; Appendix~\ref{appendix:interaction_cost}), suggesting that externalization did not introduce additional interaction cost. Post-session system-level ratings further supported these findings (Figure~\ref{fig:results}c): all 16 participants rated the system as helpful (100\%), and the majority found it fun (88\%), felt in control (88\%), and reported that it encouraged exploration (81\%). Most indicated they would use such a system (88\%) and that it kept them engaged during viewing (88\%) (Figure~\ref{fig:results}d).

\vspace{-1em}
\rev{
\section{Discussion}
\subsection{Design Implications}
}
\rev{
Our main empirical insight is the visibility-repair gap: making interpretations visible did not reliably lead to correction. Beyond the evaluated cases, our findings suggest three broader principles for externalizing AI interpretations in situated visualization systems.
}

\textbf{Externalization as a grounding mechanism, not only error recovery.} The strongest behavioral effect appeared in aligned trials, where referential externalization was accompanied by active verification even when no error was present, consistent with Clark and Brennan's grounding framework~\cite{clark1991grounding}. This contrasts with the prevailing XAI paradigm where explanations serve primarily as diagnostic tools after errors~\cite{chromik2021human, xu2023xair}. While this effect reached significance only for referential queries, the consistent directional trend across all four types suggests that externalization may function as a continuous grounding channel in situated, real-time contexts. Combined with no measurable interaction overhead, this supports exploring externalization as a persistent default rather than an opt-in feature.

% \textbf{Correction affordance as a distinct design axis from transparency.} The 62\% visibility-action gap challenges the assumption that visibility naturally leads to corrective action~\cite{bucinca2021trust, nourani2021anchoring}. The binding constraint was not externalization quality but correction cost, which varied by ambiguity type. While the absolute repair rates were low (5.6\% under externalization), the significant overall increase and per-type patterns indicate that externalization shifted users' tendency to correct, but speech-only interaction imposed a cost floor. Pending replication with larger per-type samples, future designs could explore correction modalities matched to each ambiguity type: direct selection for referential, gestural scrubbing for temporal, and toggle-based scope adjustment for metric. The guiding principle is that each modality operates on the same representation externalization already makes visible, reducing the path from noticing to acting. This gap thus reflects not a failure of externalization but a property likely shared across speech-driven situated systems: mechanisms for making interpretations transparent and mechanisms for enabling correction must be designed as separate, complementary layers. Verification beyond soccer viewing is needed to confirm this generalization.

\textbf{Correction affordance as a distinct design axis from transparency.} The $62\%$ visibility-action gap challenges the assumption that visibility naturally leads to corrective action~\cite{bucinca2021trust, nourani2021anchoring}. The binding constraint was correction cost, not externalization quality, and this cost varied by ambiguity type. Although absolute repair rates were low ($5.6\%$ under externalization), the significant overall increase and per-type patterns indicate that externalization shifted users' tendency to correct, while speech-only interaction imposed a cost floor. 
% Pending replication with larger per-type samples, future designs could explore ambiguity-specific correction: direct selection for referential, gestural scrubbing for temporal, and toggle-based scope adjustment for metric. 
% The guiding principle is to operate on the same representation externalization already makes visible, shortening the path from noticing to acting. 
\rev{Correction should reuse each visible cue to reduce verbal repair cost: tap referential candidates, scrub temporal windows, toggle metric scopes, or select spatial zones when precision matters. This shortens the path from noticing to acting.}
Thus, making interpretations transparent and enabling correction should be designed as separate, complementary layers, with verification beyond soccer viewing needed to test this generalization.

\textbf{Not all interpretation gaps require correction.} For spatial queries, externalization did not increase repair, likely because zone overlays provided close-enough interpretations that users accepted. While our spatial results are based on limited trials, this is consistent with Setlur et al.'s finding that users accept reasonable defaults for underspecified queries~\cite{setlur2019inferencing}. This suggests distinguishing between ambiguity types where precision matters (e.g., referential, where the wrong player yields a wrong answer) and those where approximate interpretations may suffice.
\rev{Thus, systems should prioritize low-cost correction for high-consequence errors, such as wrong referents or replay windows, while allowing approximate defaults for lower-stakes spatial cases.}

\vspace{1em}
% Externalization design need not be uniform: investment in correction mechanisms should adapt to how consequential a misinterpretation would be for each type.

\subsection{Limitations}

\noindent\textbf{Visual richness confound.}
Because we compare externalized interpretation with voice-only interaction, the observed benefits may partly reflect added visual information rather than externalization itself. As discussed in Section~\ref{sec:conditions}, we chose this baseline to reflect the current status quo and to follow established XAI evaluation practices~\cite{rong2024human}, but it does not isolate visual richness. Still, our results do not fit a simple ``more visuals help'' account: repair occurred in only $38\%$ of misaligned externalization trials, confirmation time did not increase, and inspectability improved selectively. \rev{Thus, we evaluate visible interpretation cues over answer-only interaction, but do not test whether tailored designs outperform labels, candidate lists, or generic highlights.}

% \textbf{Visual richness confound.} The comparison between externalized interpretation and voice-only interaction means that observed benefits may partly reflect additional visual information rather than externalization per se. As discussed in Section~\ref{sec:conditions}, we chose this baseline to reflect the current status quo and to follow established XAI evaluation practices~\cite{rong2024human}, though this design does not isolate the contribution of visual richness. However, several patterns in our data are difficult to explain by visual richness alone: referential externalization prompted increased repair in aligned trials where additional information should reduce rather than increase correction attempts, and the 62\% visibility-action gap shows that surfaced assumptions did not reliably lead to corrective action. Moreover, confirmation time did not differ between conditions, and inspectability improved selectively (three of four items, not uniformly), both inconsistent with a blanket visual-richness account. Isolating visual richness from interpretation content remains an important direction for future work.
% \rev{Thus, our study evaluates visible interpretation cues over answer-only interaction, but does not test whether tailored designs outperform simpler labels, lists, or highlights.}

\textbf{Induced misalignment and speech-only correction.} Misalignment was induced through prompt manipulation rather than arising naturally, ensuring balanced trials but potentially narrowing the range of interpretation errors. The system routes all correction through speech, which imposes a verbal construction cost that externalization alone cannot overcome. Additionally, our keyword-based repair classifier captures only explicit correction language and may undercount implicit strategies, meaning the true repair rate may be higher than reported. \rev{Finally, multimodal repair through pointing, gaze, or tapping could reduce this cost for deictic references such as ``that player'' or ``over there.''}

\textbf{LLM interpretation consistency.} The system uses GPT-4o for both ambiguity classification and interpretation, which may produce different outputs for identical queries across sessions. While classification reliability was evaluated offline (Section~\ref{sec:implementation}), we did not measure within-session consistency of the full interpretation pipeline. Variability in LLM outputs could affect the reproducibility of externalized interpretations across repeated interactions. 
% \rev{Misclassifying ambiguity type may show the wrong cue, making an incorrect interpretation seem plausible.}
\rev{Because the detected ambiguity type selects the cue, misclassification may show the wrong cue and make an incorrect interpretation seem plausible.}

\rev{\textbf{Domain and sport specificity.} Our formative study, design space, and evaluation are grounded in soccer viewing. Although the ambiguity types and externalization framework may transfer to other sports or domains, visual design, and repair costs are likely domain-dependent. Generalization would require sport-specific tracking/event data, entity identities, and mappings from ambiguities to visual cues and repair actions. Evaluation beyond soccer is needed to test this transfer.}

% \textbf{Ecological validity and sample size.} The study used pre-recorded match data as short looping clips and a 5-minute segment.
\rev{\textbf{Study scope.} Pre-recorded clips and $N=16$, consistent with within-subjects XR studies~\cite{lin2022omnioculars, chen2023iball}, supported experimental control but limit live-viewing and per-type generalizability for spatial queries.}

% \vspace{-1.0em}
\section{Conclusion and Future Work}

We investigated how externalizing a system's interpretation of ambiguous spoken queries supports inspection and correction in XR sports viewing. Externalization was associated with higher inspectability and increased repair language without added interaction cost, yet revealed a gap between visibility and corrective action that varied by ambiguity type, suggesting that correction affordances must be designed independently from transparency. This visibility-action gap may extend to other speech-driven situated systems, though evaluation beyond soccer viewing is needed. Future work should explore multimodal correction mechanisms matched to each ambiguity type~\cite{lee2021whatsthis, lee2024gazepointar} and evaluation with live broadcasts and other sports domains.

\acknowledgments{%
	This work was supported by NIH grant 1U01CA284207 and Harvard Data Science Initiative Trust in Science Fund Award.
}

\raggedbottom
\bibliographystyle{abbrv-doi-hyperref}

\bibliography{template}

\clearpage
\appendix
% ============================================================
% Supplementary Material: Formative Study
% ============================================================
 
\section{Supplementary Material}

% ------------------------------------------------------------
% A. XR Sports App Survey
% ------------------------------------------------------------
\subsection{Survey of Existing XR Sports Viewing Applications}
\label{sec:suppl_apps}

To contextualize our design, we surveyed nine existing XR sports viewing  applications across platforms including Apple Vision Pro and Meta Quest. Table~\ref{tab:xr_apps} summarizes the spatial layout and information  types provided by each application. Across the surveyed apps, we observed a consistent spatial pattern:  the \textit{center} region is almost universally reserved for the main  video feed, serving as the primary viewing anchor. Supplementary information (e.g., player statistics, game scores, and rankings) is distributed across \textit{left} and \textit{right} panels, while several apps utilize a \textit{tabletop} surface for additional contextual data (e.g., miniature field maps or mixed overlays). Notably, most apps present a fixed set of predefined information panels. Users cannot dynamically request new data or specify what they want to see. This survey reveals two key gaps relevant to our work. First, none of the surveyed apps support \textit{speech-driven interaction} for querying game information. Users must navigate through pre-arranged panels rather than asking questions in natural language.  Second, the information presented is \textit{statically determined}  by the app rather than adapted to the user's current intent or viewing context. As a result, when users have specific, context-dependent questions (e.g., about a particular player's recent performance or the reason behind a referee's decision),  existing apps offer no mechanism for resolving such queries, let alone handling the ambiguity inherent in natural language requests. These observations motivated our exploration of speech-driven sports viewing, where the system must interpret and disambiguate user utterances in real time.

\begin{table}[h]
\centering
\caption{Survey of existing XR sports viewing applications. Layout columns indicate what content is placed in each spatial region. ``O'' indicates the region is used; ``X'' indicates it is absent.}
\label{tab:xr_apps}
\resizebox{\columnwidth}{!}{%
\begin{tabular}{llllll}
\toprule
\textbf{App} & \textbf{Left} & \textbf{Center} & \textbf{Tabletop} & \textbf{Right} & \textbf{Sport} \\
\midrule
Lapz & --- & Main Video & O & Additional Video & Car Racing \\
ESPN & Current Game & Main Video & Mixed & Player Info & Basketball \\
NASCAR & Player Stats & Main Video & O & Ranking & Car Racing \\
SailGP & Stats / Map & Main Video & O & Video / Ship & Sail Racing \\
MLB & Current Game & Main Video & O & Player Stats & Baseball \\
PGA Tour & X & Game Stats & O & X & Golf \\
Xtadium & Video & Main Video & X & Video & Multi-Sports \\
Vroom & X & Main Video & O & Current Game & Car Racing \\
Immersiv & Players & Main Video & O & Current Game & Soccer \\
\bottomrule
\end{tabular}%
}
\end{table}
 
% ------------------------------------------------------------
% B. Guided Prompts
% ------------------------------------------------------------
\subsection{Scenario Descriptions for Question Elicitation in Formative Study}
\label{sec:suppl_prompts}
 
During the question-elicitation phase of the formative study, participants watched five short video clips (approximately 30 seconds each) and one longer video segment (approximately 4 minutes).
For each short clip, we provided two scenario-based prompts grounded in common question categories from prior sports-viewing research~\cite{lin2022omnioculars, lee2024sportify}, followed by a free-question period.
Table~\ref{tab:prompts} lists the full set of scenario descriptions used in the formative study.

\begin{table}[h]
\centering
\caption{Guided prompts provided for each video clip during the formative study.}
\label{tab:prompts}
\resizebox{\columnwidth}{!}{%
\begin{tabular}{clll}
\toprule
\textbf{Clip} & \textbf{Scenario} & \textbf{Prompt} & \textbf{Category} \\
\midrule
\multirow{2}{*}{1} & Goal scored & ``You know who scored and want to pull up the scorer's detailed profile.'' & Identification \\
 & Comparison & ``You're curious how the goal scorer's performance compares to the other attackers.'' & Comparison \\
\midrule
\multirow{2}{*}{2} & Card shown & ``A card was just shown and you're not sure exactly what led to it.'' & Event Description \\
 & Foul review & ``You want to see the foul situation again because you're not sure the card was deserved.'' & Nav. / Replay \\
\midrule
\multirow{2}{*}{3} & Shot on post & ``A shot just hit the post and you want to know which player took it.'' & Identification \\
 & Shooting stats & ``You want to check the shooting stats.'' & Stat. Retrieval \\
\midrule
\multirow{2}{*}{4} & Turnover & ``You want to know who made the turnover or missed pass.'' & Identification \\
 & Passing stats & ``You want to see the passing stats.'' & Stat. Retrieval \\
\midrule
\multirow{2}{*}{5} & Walkthrough & ``You want the system to walk you through what happened.'' & Event Description \\
 & Goal replay & ``You want to see the goal moments again.'' & Nav. / Replay \\
\midrule
Long & \multicolumn{3}{l}{\textit{No guided prompts; participants asked questions freely throughout.}} \\
\bottomrule
\end{tabular}%
}
\end{table}

\subsection{Scenario Descriptions for Main User Study}
\label{sec:main_study_prompts}

During the main user study, participants completed eight video clip scenarios (A1--A4 from Match~A: 1.~FC~K\"oln vs.\ FC~Bayern M\"unchen; B1--B4 from Match~B: VfL~Bochum~1848 vs.\ Bayer~04~Leverkusen). Each scenario contained two sequential descriptions that oriented participants toward match-relevant queries without prescribing specific wording. Table~\ref{tab:main_prompts} lists the full set of guided prompts used in the study.

\begin{table}[h]
\centering
\caption{Guided prompts for each scenario in the main user study.}
\label{tab:main_prompts}
\scriptsize
\begin{tabular}{cp{6.2cm}}
\toprule
 & Task Description \\
\midrule
A1-1 & ``Bayern just scored through a slick passing move. Multiple players were involved --- you might want to ask about the attack.'' \\
A1-2 & ``The attack had several phases. You could ask about specific moments in the sequence.'' \\
\midrule
A2-1 & ``Bayern are creating pressure with multiple events in quick succession. Try asking about when something specific happened.'' \\
A2-2 & ``A dangerous chance was created. You might want to ask about the numbers behind it.'' \\
\midrule
A3-1 & ``A Bayern player dribbles past opponents before getting fouled. Something stands out about his performance.'' \\
A3-2 & ``The foul happens somewhere specific on the field. You could ask about the location.'' \\
\midrule
A4-1 & ``Bayern take a corner kick and the ball bounces around the penalty area. There's a lot going on in different parts of the field.'' \\
A4-2 & ``Multiple Bayern players contested the ball in the box. Someone made a big impact.'' \\
\midrule
B1-1 & ``Two players get involved in a physical challenge. It's not obvious who did what at first glance.'' \\
B1-2 & ``There were multiple physical duels during this sequence. You might wonder when the key one happened.'' \\
\midrule
B2-1 & ``Bochum build toward a goal through several passes and a cross. Multiple moments stood out in the sequence.'' \\
B2-2 & ``A goal is scored. You might want to know more about the goalscorer's performance in numbers.'' \\
\midrule
B3-1 & ``A goal is scored from a corner kick delivery. Several players were involved in different roles.'' \\
B3-2 & ``The goal started from a corner. The action moved through different parts of the field.'' \\
\midrule
B4-1 & ``Leverkusen lose possession suddenly and Bochum break forward quickly. Something on the field triggered the whole counter-attack.'' \\
B4-2 & ``Multiple Bochum players are involved in the counter-attack. Someone ends up making the biggest difference.'' \\
\bottomrule
\end{tabular}
\end{table}

\subsection{System Prompts}
 
This document provides the full system prompts used in our XR soccer viewing system. Section~\ref{sec:base} presents the base prompt used for ambiguity-aware interpretation and response generation during interactive use.  Section~\ref{sec:misalignment} presents the misalignment induction prompt appended to generate deliberately shifted interpretations for misaligned trials (Section~6.1).
 
This prompt is sent to GPT-4o for every user query during interactive use. It defines the JSON response format, ambiguity type definitions, candidate structure, visualization types, and rules for interpreting real-time match state.

\subsection{Misalignment Induction Prompt}
\label{sec:misalignment}
 
The following prompt was appended to the base prompt during \textit{misaligned trials} in both conditions (Section~6.1). Following prior work that deliberately manipulates system outputs to study detection and repair behavior~\cite{lakkaraju2020fool, heuer2020accuracy}, this override instructs the model to introduce a subtle factual error (e.g., selecting a plausible but non-primary player, shifting a timestamp, or choosing a secondary metric), ensuring that misaligned interpretations remained within the space of plausible readings.
 
\begin{lstlisting}
NOTE: Answer the question but intentionally
introduce a small factual error -- pick a
plausible but wrong player, wrong time, or wrong
stat. Sound confident. The error should be subtle,
not obvious.
\end{lstlisting}
 
In practice, scenario-specific constraints were also appended alongside this override to target particular ambiguity types. For example, in a referential scenario where Bayern scored through a passing sequence, the constraint directed the model to select a different but plausible passer from earlier in the sequence rather than the most salient one.

\section{Statistical Results}
\label{appendix:stats}
 
Tables~\ref{tab:subjective}--\ref{tab:interaction_cost} report the full Wilcoxon signed-rank test results for all pairwise comparisons in the user study. All tests were conducted on per-participant paired means ($N=16$), and effect sizes are reported as $r = |Z|/\sqrt{N}$. Significant results ($p < .05$) are marked with $*$.

\subsection{Subjective Ratings (Sec 7.1)}
\label{appendix:subjective}
 
Table~\ref{tab:subjective} reports pairwise comparisons for all subjective rating items. Externalization significantly improved three of four inspectability items and overall satisfaction. Participants found it easier to identify which player the system referred to ($p = .011$, $r = .62$), which moment was shown ($p = .003$, $r = .73$), and what alternatives were considered ($p = .024$, $r = .56$). Identifying what went wrong showed a numerical advantage for externalization but did not reach significance ($p = .151$, $r = .36$), suggesting that externalization may support recognizing what the system assumed more readily than diagnosing why it erred. Overall satisfaction was significantly higher under externalization ($p = .002$, $r = .73$). Trust and perceived clutter did not differ significantly between conditions, indicating that the observed benefits were perceptual and experiential rather than reflecting a generalized shift in system confidence or an attentional burden.
 
\begin{table}[h]
\centering
\caption{Subjective rating comparisons (7-point Likert) between baseline and externalization conditions.}
\label{tab:subjective}
\footnotesize
\begin{tabular}{lcccccc}
\toprule
Item & Base $M$ & Ext $M$ & $W$ & $p$ & $r$ \\
\midrule
Which player referred to & 4.69 & 5.94 & 3.0 & .011$^*$ & .62 \\
Which moment shown & 4.69 & 6.25 & 4.0 & .003$^*$ & .73 \\
Alternatives considered & 3.62 & 4.75 & 5.5 & .024$^*$ & .56 \\
What went wrong & 4.31 & 5.06 & 10.5 & .151 & .36 \\
Overall satisfaction & 4.56 & 5.69 & 0.0 & .002$^*$ & .73 \\
Trust & 4.44 & 5.00 & 17.5 & .297 & .25 \\
Perceived clutter & 3.25 & 3.56 & 22.5 & .607 & .13 \\
\bottomrule
\end{tabular}
\end{table}

\subsection{Overall Repair Rate and Manipulation Check (Sec 7.2)}
\label{appendix:repair_overall}
 
Table~\ref{tab:repair_overall} reports the overall repair utterance rate comparison and the manipulation check comparing follow-up query counts between aligned and misaligned trials. Repair utterance rate was computed as the proportion of all queries within a condition that contained at least one of the 13 keyword patterns described in Section~6.3. The significant difference between conditions ($p = .039$, $r = .57$) indicates that externalization shifted participants toward more explicit correction language. As a manipulation check, we compared the total number of follow-up queries participants produced in aligned versus misaligned trials, collapsing across conditions. Participants produced significantly more follow-up queries in misaligned trials ($M_{\text{mis}} = 33.75$) than in aligned trials ($M_{\text{align}} = 27.62$; $W = 13.0$, $p = .004$), confirming that the induced misalignment produced observable behavioral differences and that participants were sensitive to interpretation errors.

\begin{table}[h]
\centering
\caption{Overall repair rate and manipulation check.}
\label{tab:repair_overall}
\begin{tabular}{lccccc}
\toprule
\multicolumn{6}{l}{\textit{Baseline vs.\ Externalization}} \\
\midrule
Measure & $M_{\text{base}}$ & $M_{\text{ext}}$ & $W$ & $p$ & $r$ \\
\midrule
Repair rate & 3.3\% & 5.6\% & 16.0 & .039$^*$ & .57 \\
\midrule
\multicolumn{6}{l}{\textit{Aligned vs.\ Misaligned}} \\
\midrule
Measure & $M_{\text{align}}$ & $M_{\text{misalign}}$ & $W$ & $p$ & $r$ \\
\midrule
Follow-up queries & 27.62 & 33.75 & 13.0 & .004$^*$ & --- \\
\bottomrule
\end{tabular}
\end{table}
 
% \begin{table}[t]
% \centering
% \caption{Overall repair rate and manipulation check.}
% \label{tab:repair_overall}
% \begin{tabular}{lcccccc}
% \toprule
% Measure & Condition 1 & Condition 2 & $W$ & $p$ & $r$ \\
% \midrule
% Repair rate & Baseline: 3.3\% ($SD$=4.2\%) & Ext.: 5.6\% ($SD$=3.8\%) & 16.0 & .039$^*$ & .57 \\
% Follow-up queries & Aligned: 27.62 & Misaligned: 33.75 & 13.0 & .004$^*$ & --- \\
% \bottomrule
% \end{tabular}
% \end{table}
 
\subsection{Per-Type Repair Rate in Aligned Trials (Sec 7.3)}
\label{appendix:repair_aligned}
 
Table~\ref{tab:repair_aligned} reports per-type repair rate comparisons in aligned trials. Given the small per-cell sample sizes from crossing ambiguity type with alignment condition, these analyses are exploratory and should be interpreted as indicative patterns rather than confirmatory findings. Only referential queries showed a significant increase in repair rate under externalization ($W = 0.0$, $p = .018$, $r = .89$), where participants produced repair language in 14.1\% of externalization trials compared to 0.0\% under baseline. This pattern is consistent with the grounding account discussed in Section~7.3: visible candidate badges provided a concrete commitment against which participants could verify their own intent, prompting active inspection even when the system's interpretation was correct. The remaining three ambiguity types showed no significant differences, though the numerical trend under externalization was equal to or higher than baseline across all four types.

\begin{table}[h]
\centering
\caption{Per-type repair rate (\%) in aligned trials (baseline vs.\ externalization).}
\label{tab:repair_aligned}
\begin{tabular}{lcccccc}
\toprule
Ambiguity Type & Baseline $M$ & Ext.\ $M$ & $W$ & $p$ & $r$ \\
\midrule
Referential & 0.0\% & 14.1\% & 0.0 & .018$^*$ & .89 \\
Temporal & 0.0\% & 3.8\% & 0.0 & .180 & --- \\
Metric & 1.6\% & 1.7\% & 3.0 & 1.000 & --- \\
Spatial & 1.2\% & 3.1\% & 1.0 & .655 & --- \\
\bottomrule
\end{tabular}
\end{table}
 
\subsection{Per-Type Repair Rate in Misaligned Trials (Sec 7.4)}
\label{appendix:repair_misaligned}
 
Table~\ref{tab:repair_misaligned} reports per-type repair rate comparisons in misaligned trials. None of the four ambiguity types showed a significant increase in repair rate under externalization. Referential and spatial repair rates remained similar across conditions (both $p > .5$), suggesting that for these types, externalization did not lower the threshold for corrective action when misinterpretation was present. Temporal and metric queries showed large effect sizes (temporal $r = .64$; metric $r = .91$) but did not reach significance ($p = .194$; $p = .066$), likely reflecting limited statistical power from small per-cell sample sizes rather than absence of an effect. We treat these as suggestive patterns warranting replication with larger samples. The overall pattern contributes to the 62\% visibility-action gap discussed in Section~7.4, where externalization surfaced system assumptions but participants often did not act on them, particularly when correction required constructing precise verbal specifications.

\begin{table}[h]
\centering
\caption{Per-type repair rate (\%) in misaligned trials (baseline vs.\ externalization).}
\label{tab:repair_misaligned}
\begin{tabular}{lcccccc}
\toprule
Ambiguity Type & Baseline $M$ & Ext.\ $M$ & $W$ & $p$ & $r$ \\
\midrule
Referential & 11.2\% & 9.9\% & 11.0 & .611 & .19 \\
Temporal & 1.2\% & 7.5\% & 1.5 & .194 & .64 \\
Metric & 0.0\% & 7.4\% & 0.0 & .066 & .91 \\
Spatial & 6.1\% & 3.0\% & 6.0 & .686 & .18 \\
\bottomrule
\end{tabular}
\end{table}
% \subsection{Guided Prompts for Question Elicitation}
% \label{sec:suppl_prompts}
% Repair utterances were identified using a keyword-based classifier organized into four categories derived from interview responses and manual inspection of query texts. \textit{Explicit correction} markers indicate direct self-correction (``I mean,'' ``I meant,'' ``meant,'' ``not that,'' ``actually''). \textit{Negation and redirection} markers signal rejection of the system's interpretation or reference to an alternative (``no,'' ``the other one,'' ``what about the other,'' ``wrong,'' ``different,'' ``other,'' ``instead,'' ``that's not''). \textit{System challenge} markers indicate the user questioning the system's response (``are you sure,'' ``I don't think,'' ``really?,'' ``isn't it''). \textit{Re-request and clarification} markers indicate attempts to re-specify the original intent (``I was asking,'' ``supposed to,'' ``more specific,'' ``not that play''). Keywords were matched longest-first to avoid overlap. 

\subsection{Interaction Cost (Sec 7.5)}
\label{appendix:interaction_cost}
 
Table~\ref{tab:interaction_cost} reports pairwise comparisons for three interaction cost metrics between the baseline and externalization conditions. Query count per scenario measured the total number of queries participants issued within each scenario, capturing overall interaction volume. Confirmation time measured the interval in seconds from the system's first response to the participant's next query, serving as a proxy for how long participants spent evaluating each response before continuing. Query word count measured the average number of words per query, capturing whether externalization led participants to produce longer or more detailed utterances. None of the three comparisons reached statistical significance (all $p > .05$), indicating that externalization did not impose additional interaction overhead in terms of query frequency, response evaluation time, or utterance complexity.
 
\begin{table}[h]
\centering
\caption{Interaction cost comparisons between baseline and externalization conditions. None reached significance.}
\label{tab:interaction_cost}
\begin{tabular}{lcccccc}
\toprule
Measure & Baseline $M$ & Ext.\ $M$ & $W$ & $p$ & $r$ \\
\midrule
Query count & 7.45 & 7.89 & 60.0 & .679 & .10 \\
Confirmation time (s) & 25.8 & 25.9 & 68.0 & 1.000 & .00 \\
Query word count & 8.0 & 8.0 & 63.0 & .821 & .06 \\
\bottomrule
\end{tabular}
\end{table}

\clearpage
\onecolumn
\section{System Prompt}

\label{sec:base}
\begin{lstlisting}
You are a sports XR assistant for live Bundesliga soccer match visualization (2022-23 season). You receive a voice transcript, real-time match state, and reference data.
 
Respond with ONLY a JSON object (no markdown, no extra text):
{
  "assistant_text": "your English answer here",
  "ambiguity_type": "referential|spatial|temporal|metric",
  "confidence": 0.0-1.0,
  "candidates": [{"id": "B_05", "label": "PlayerName", "confidence": 0.8, "role": "passer"}, {"id": "B_07", "label": "OtherPlayer", "confidence": 0.5, "role": "receiver"}],
  "evidence": ["reason1", "reason2"],
  "ambiguity_reasons": ["3-4 word reason"],
  "visualizations": [{"type": "highlight",  "target": "B_05"}, {"type": "heatmap", "target": "R_03"}],
  "metrics_context": {"count_type": "goals", "temporal_ref": "this_match","metric_type": "possession"}
}
 
Ambiguity types:
- "referential": WHO is ambiguous. The intended person/group is underspecified -- pronouns ("he", "that player"), role descriptions ("the midfielders", "the player who scored"), or unnamed entities have multiple candidates. CRITICAL: the "candidates" array MUST have exactly 2 or 3 entries. NEVER return just 1 candidate for referential ambiguity. Each candidate needs "id" (player ID like B_05), "label" (real player name), "confidence" (0.0-1.0), and "role" (their role in the action). The first candidate must be the most likely answer and MUST match the player mentioned in assistant_text and visualizations[0].target. "role" values for referential: "passer", "receiver", "shooter", "fouled", "defender", "assist", "tackler", "crosser", "keeper", "candidate", "blocker", "marker", "interceptor", "dribbler", "scorer", "creator", "runner", "clearance" Each candidate MUST also include "reason": a free-text phrase (max 5 words) explaining why this candidate is relevant to the action. Examples: "cross", "key pass", "failed to interrupt", "lost the runner", "failed to intercept", "closest to ball", "involved in buildup", "missed the tackle", "blocked the shot", "late to cover", "overlapping run".
 
- "spatial": WHERE is ambiguous. Each candidate "role" must be a SPECIFIC zone name from this list: Small zones: "left-upper", "left-lower", "right-upper", "right-lower", "center-left", "center-right", "penalty area", "home penalty area", "corner area" Medium zones: "left wing", "right wing", "center", "midfield" Large zones (use only when the question is about a whole side): "left third", "right third", "defensive third", "attacking third" Prefer SMALL zones. Only use medium/large when the question explicitly asks about a broad area. ambiguity_reasons: EXACTLY 1 item, pick from: "zone boundary", "overlapping areas", "vague reference", "multiple zones" Each spatial candidate MUST also include: "reason": free-text phrase (max 6 words) explaining why this zone was selected. Examples: "shot attempt here", "crossing point", "gap on the side", "space near the goal", "ball location at event", "low player density", "open passing lane", "gap between defensive lines", "near recent foul", "space behind fullback". "rank": integer indicating relevance order (1 = most likely zone). Rank must be consistent with confidence.
 
- "metric": WHAT METRIC / WHICH STAT is ambiguous. The user asks about stats,  performance, rankings, counts, rates, records, or comparisons without specifying the exact  metric, timeframe, or comparison baseline. This is a BROAD category -- it includes:  
  * Direct stat questions: "How many passes?", "What's the success rate?"
  * Performance evaluations: "Is he good?", "Is he getting better?", "doing well?"
  * Rankings/standings: "What is the ranking?","league position?"
  * Season/career records: "How is the season going?", "scoring records"
  * Requests for statistical data: "Tell me statistical data", "show me the numbers"
  * Implicit evaluations: "He's good at heading" (WHICH metric quantifies this?)
  * Ability/skill questions: "Is there a player good at X?" (measured HOW?)
  If the query involves ANY performance, record, ranking, ability evaluation, or statistical information -- even phrased indirectly -- classify as metric. 
  Each candidate "role": "pass stat", "shot stat", "xG stat", etc.
  ambiguity_reasons: EXACTLY 1 item, pick from: "which stat?", "which timeframe?", "which metric?"
  MUST include "metrics_context" with:
    count_type: "goals"|"shots"|"fouls"|"passes"|"assists"
    temporal_ref: "this_play"|"this_match"|"this_season"|"last_season"|"last_5_games"
    metric_type: "possession"|"xG" (or null)
 
- "temporal": WHEN is ambiguous -- user asks about timing, sequence, or "what happened" across multiple events. Use for: "what happened in that sequence?", "when did he score?", "show me the key moments", "what just happened?" candidates: list the events as candidates.
  Each candidate needs: 
    "id": player ID involved "label": event description (e.g. "Tackle by Kimmich")
    "role": event type ("tackle", "shot", "goal", "pass", "foul", "cross", "save", "clearance", "interception", "corner", "free_kick", "card")
    "confidence": 0.0-1.0
    "timestamp": match time string (e.g. "18:23")
    "description": brief contextual phrase (max 4 words). Examples:
      "started the chance", "delivered the ball", "finished the attack", "key defensive action", "set piece delivery", "cross into the box"
    "rank": integer for chronological position (1 = earliest)
  The candidate with highest confidence is the system's selected moment.
  evidence: list event types for timeline filtering, e.g. ["tackle", "shot", "goal"].
  ambiguity_reasons: 
  EXACTLY 1 item, pick from: "multiple events", "close timing", "unclear reference", "repeated pattern"
 
- "none": ONLY use when the query is fully unambiguous, OR is a general rule/knowledge question ("What does a yellow card mean?"), OR requires visual/external info not in match data ("Why did he get a card?" -- needs video). No ambiguity_reasons needed. Do NOT classify stat/metric questions as none -- those are metric.
 
IMPORTANT: ambiguity_reasons must be EXACTLY 1 item, max 3 words. Never more than 1 reason.
 
IMPORTANT: Most user queries in XR contain SOME ambiguity. Prefer referential/temporal/ metric/spatial over none when applicable.
 
Player IDs: Each player has a unique spawn ID (e.g. B_GK, B_00~B_09 for home, R_GK, R_00~R_09 for away). These IDs are provided in the UI State roster under each player's "id" field and in clip_events as "player_id". ALWAYS use these exact IDs in candidates and visualizations -- do NOT guess or make up IDs.
 
CRITICAL: Look at the "team" field in clip_events to determine which team performed the action. B_ prefix = home team, R_ prefix = away team. If a pass event shows team="home", the passer MUST be a B_ player. If team="away", it MUST be an R_ player. NEVER mix up home/away teams.
 
Set confidence to how sure you are (0.0=guess, 1.0=certain).
 
Rules:
- ALWAYS respond in English regardless of the input language.
- Keep assistant_text concise: 1 sentence, HARD LIMIT 60 characters. Be direct -- name the player, action, or stat. No padding.
- No markdown code fences.
- Use the provided data to give accurate, specific answers.
- Always include ambiguity_type and confidence. Candidates/evidence optional if type is "none".
- When ambiguity_type is NOT "none", the "candidates" array MUST have at least 2 entries (never just 1). Each candidate needs id, label, and confidence.
- Use conversation history to maintain context. If the user says "he" or "that player", refer to the previously discussed player.
- ALWAYS include "visualizations" array with 1-3 items to show relevant data on the 3D field. This is REQUIRED for every response.
  Available types:
  * "highlight" -- highlight a player (target = player ID)
  * "heatmap" -- show player movement heatmap (target = player ID)
  * "trajectory" -- show ball/player path (target = player ID or "Ball")
  * "formation" -- show team formation (target = "B" or "R")
  * "speed_label" -- show speed above player (target = player ID)
  * "zone" -- highlight field zone (target = comma-separated player IDs)
  * "relation_link" -- show passing chain (target = comma-separated player IDs)
- At minimum, highlight the player(s) being discussed. Never return an empty visualizations array.
 
CRITICAL for time-sensitive questions
("who passed?", "who has the ball?", etc.):
- clip_events are in chronological order. The LAST event is the most recent.
- Always prioritize the MOST RECENT relevant event. For "who passed the ball?", use the "last_pass" field directly.
- Play_Cross is a type of pass (a cross into the box). Treat it the same as Play_Pass when answering pass-related questions.
- NEVER pick an older pass event when a newer one exists. The "last_pass" field is authoritative.
 
Data you receive:
1. [UI State] - Real-time match context: match_time, half, score, possession, clip_events (chronological order, last = most recent, with player_id = spawn ID), team rosters (each player has name, num, pos, and id = spawn ID)
2. [Reference Data] - Relevant context selected
   per question:
   - Match Detail: lineups, goals, cards, substitutions, fouls, set pieces, individual player stats, match ratings
   - Player Profiles: bio, nationality, height, position, career stats, scouting notes, debut info, youth career
   - Team Profiles: club history, stadium, titles, notable players, current season performance
   - League Tables: Bundesliga standings (2020-21, 2021-22, 2022-23)
   - Top Scorers: season goal and assist leaders
   - Squad Stats: per-player season stats (apps, goals, assists, cards) across multiple seasons
   - Comparison Players: external players (Son Heung-min, Haaland, Bellingham, Nkunku, Lewandowski) for cross-comparison
 
Answer naturally about: match events, player identification, stats, comparisons (including external players), league standings, team/player history, fouls/cards, set pieces, substitutions, and background context.
\end{lstlisting}

% \appendix % You can use the `hideappendix` class option to skip everything after \appendix

\end{document}